\documentclass[fleqn,10pt]{wlscirep}
\usepackage[utf8]{inputenc}
\usepackage[T1]{fontenc}
\usepackage{here}
\usepackage{booktabs}

\title{Gender differences in collaboration and career progression in physics}

\author[1,*]{Mingrong She}
\author[2,3,4]{Jan Bachmann}
\author[2,3]{Fariba Karimi}
\author[1]{Leto Peel}
\affil[1]{Department of Data Analytics and Digitalisation, School of Business and Economics, Maastricht University, Maastricht, The Netherlands}
\affil[2]{Graz University of Technology, 8010 Graz, Austria}
\affil[3]{Complexity Science Hub, Vienna, Austria}
\affil[4]{Department of Network and Data Science, Central European University, Vienna, Austria}

\affil[*]{m.she@maastrichtuniversity.nl}

\begin{abstract}
We examine gender differences in collaboration networks and academic career progression in physics. We use the likelihood and time to become a principal investigator (PI) and the length of an author's career to measure career progression. Utilising logistic regression and accelerated failure time models, we examine whether the effect of collaboration behaviour varies by gender. We find that, controlling for the number of publications, the relationship between collaborative behaviour and career progression is almost the same for men and women. Specifically, we find that those who eventually reach principal investigator (PI) status, tend to have published with more unique collaborators. In contrast, publishing repeatedly with the same highly interconnected collaborators and/or larger number of co-authors per publication is characteristic of shorter career lengths and not attaining PI status. We observe that women tend to collaborate in more tightly connected and larger groups than men. Finally, we observe that women are less likely to attain the status of PI throughout their careers and have a lower survival probability compared to men, which calls for policies to close this crucial gap.
\end{abstract}

\begin{document}

\flushbottom
\maketitle
%
%
\thispagestyle{empty}

\section{Introduction}
The field of physics has a long history of gender disparities, in which the majority of researchers have been men.
Research has shown that women are underrepresented at all levels of physics education and in the physics workforce~\cite{Sax2016, ivie2012women}. Furthermore, female researchers in physics have been found to encounter greater barriers to professional advancement and tend to leave academia earlier than their male counterparts~\cite{Clancy1962, Ivie_2013, Holman2019}. 
A 2014 report indicates that women represent 23\% of assistant professors, 18\% of associate professors and only 10\% of full professors in physics departments in the United States~\cite{porter_women_2019}. Similar findings apply to Europe and other areas of STEM fields~\cite{ec2019she}. But what are the driving factors that help or hinder women from the career progression in physics? Here, we examine that question through the association between gender, collaboration behaviour and career progression.  

In recent years, the impact of collaboration on career progression has garnered increased attention as a potential contributing factor to these disparities~\cite{Defazio2009}. Collaboration is a crucial aspect of scientific research and has been shown to significantly impact professional development and advancement~\cite{Jessica2021collaboration,bachmann2024cumulativeadvantagebrokerageacademia}. 
A number of studies have shown that women have smaller collaboration networks than men~\cite{Cole1984ThePP, Bozeman2004Scientists, McDowell2006TWO,jadidi2018gender}. This difference in collaboration behaviour may negatively impact the career advancement of women, as collaborations are often required for scientific success and can lead to greater visibility and credibility in the field~\cite{adams2005scientific}.
Furthermore, Li et al.~\cite{li2022untangling} found that differences in collaboration networks could largely explain gendered disparities in productivity and impact.

Women are under-represented in leadership positions in academia and industry. Ginther and Kahn~\cite{Stephen2014} found that women in scientific, engineering and medical fields are less likely to be promoted to full professor than men. Similarly, Moss-Racusin et al.~\cite{Corinne2012} found that women scientists are less likely to be hired and promoted. These findings indicate a persistent disparity in the progression of women towards leadership roles in academia and industry. Career progression is not only important for individuals themselves, but can also affect the structure of the professional hierarchy for future generations. For example, more diversity and inclusion at senior management levels has been seen to affect policies and research priorities~\cite{adams2009women}. Besides, peer mentoring with individuals of the same gender during crucial developmental stages enhances the success and retention of women, producing long-term benefits, especially in more male-dominated fields such as science and engineering~\cite{stout2011steming, dennehy2017female}. 

\begin{figure}[H]
    \vspace{-7pt}
    \centering
    \includegraphics{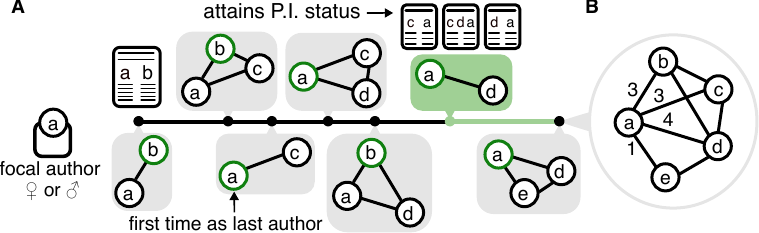}
    \caption{\textbf{Illustration of career progression and collaboration.}
    (\textbf{A}) The career trajectory of a focal author $a$, visualised as a timeline of publication events. Each solid black dot on the timeline represents a publication in time. Below each dot, the grey box illustrates the co-authorship network for a paper published at that time. Within each box, nodes represent co-authors; a green-outlined node indicates the last author on that publication. The solid green dot on the timeline marks the point at which $a$ becomes a Principal Investigator (PI), defined as having published their third last-author paper. The corresponding publication is highlighted with a green box.
    For all scientists who eventually become PI, we calculate the time to become PI as the number of years between their first publication and the publication through which they attained PI status.
    The number of years after becoming PI until the last publication define the time after PI.
    Using the careers of all focal authors, we further compute career lengths as the time passed between their first and last published papers and the likelihood of becoming a PI as the proportion of PIs among all authors.
    (\textbf{B}) The ego network of author $a$ aggregated over their entire career. Edges represent co-authorships with edge weights indicating number of shared papers. Here, author $a$ has collaborated with four unique co-authors across seven publications, yielding a network size of 4, mean tie strength of $11 / 4 = 2.75$, clustering coefficient of $4 / 6 = 0.67$ and an average of $11 / 7 = 1.57$ co-authors per paper.}
    \label{fig:fig1}
    \vspace{-7pt}
\end{figure}

In this paper, we present a comprehensive analysis of publication data from the American Physical Society (APS) that reveals several key findings (Figure~\ref{fig:fig1}). We find differences in the way men and women collaborate.
Women have stronger connections with each co-author, collaborators who are more closely connected to each other and a greater mean number co-authors per publication. We identify differences in the career progression of men and women, finding that women exhibit lower longevity than their male counterparts regardless of whether they attain a PI status. This indicates that the career progression of females in academia may not fully capture the value of these dense and collaborative networks.

\section{Selecting focal authors}
We analyse a dataset from the American Physical Society (APS) between 1893 and 2020, comprising $678,916$ publications in 19 different Physical Review journals (further details given in table~\ref{APS_data}).
Within this dataset, we use an algorithm for name disambiguation based on the method used by Sinatra et al.~\cite{sinatra2016quantifying} to identify unique authors and employ a comprehensive methodology~\cite{buskirk.etal_opensourceculturalconsensus_2023} to infer the gender of each author (see supplementary~\ref{data_process} information for full details). Using this process we identify the gender of $248,788$ unique authors. 

In order to analyse career progression in physics, we define focal authors as those researchers who meet a set of criteria related to their publication history and collaboration patterns.  We focus only on a subset of authors who have sufficient publication history to capture their collaboration behaviour and exclude authors with excessive records as these often signal incorrect merges in the name disambiguation process. 
Additionally, we assume that authors who publish in the last 5 years of the dataset (2015-2020) are potentially still active. We remove these active authors so that we only have focal authors who have ended their career or have otherwise dropped out:
\begin{enumerate}
    \item $191,300$ authors are excluded due to having published fewer than three papers or more than $400$ papers in total. 
    \item $8,708$ authors are excluded due to an interval year between first publication and last publication that is smaller than $3$ years or larger than $50$ years. 
    \item $464$ authors are excluded due to having less than two unique co-authors.
    \item $29,217$ active authors are excluded because they are still publishing after 2015. 
\end{enumerate}
After applying these filtering criteria, $19,099$ focal authors remain in the dataset, who have collectively published $191,955$ papers, including $1,486$ women and $17,613$ men.

\section{Collaboration behaviour}
To capture collaboration behaviour, we construct a weighted egocentric network, for each author, derived from joint publications with other authors (Figure~\ref{fig:fig1}). In each egocentric network, the \textit{ego} $E$ represents the focal author and \textit{alters}
are co-authors who have published papers with the focal author. The edge weight or \textit{tie strength}, indicates the number of co-authored publications. An edge or connection, exists between a pair of alters if they have co-authored at least one paper together with the ego.
This choice of representation concentrates on the focal author's direct collaborations at the cost of omitting some connections between alters that do not relate to the focal author.
We calculated four statistics for each egocentric network to measure the collaboration behaviour of the focal authors (see Figure~\ref{fig:fig1} for details): 
\begin{itemize}
    \item \textbf{Network size} is the count of unique collaborators connected to the focal author. The network size measures the collaborative reach of the focal author.  
     \item \textbf{Mean tie strength} is the sum of the edge weights in the network divided by the network size. The mean tie strength indicates the recurrence of collaborations. 
    \item \textbf{Local clustering coefficient} is the proportion of pairs of alters that are connected to each other in the network. The clustering coefficient indicates the degree to which the co-authors are interconnected among themselves. Throughout the paper, we describe ego networks with high clustering coefficients as ``tight'', reflecting strong local interconnectedness among co-authors.
    \item \textbf{Co-authors per publication} is the sum of the edge weights in the network divided by the total number of publications authored by the ego. The mean number of co-authors per publication indicates the size of the collaborations that the ego tends to engage in. 
\end{itemize}

\subsection{Adjusting for differences in number of publications}
The number of papers a focal author publishes can have an influence on the pattern of collaborations that the focal author engages in. For instance, the more a focal author publishes, the more opportunity the focal author has to increase their network size. We examine the effect of this potential confounder by setting up a linear regression to estimate an author's number of publications according to their collaboration statistics. Table~\ref{regression_papers} displays the regression coefficients that suggest a significant association between publication count and collaboration statistics. 
To address this confounding factor and allow us to more closely examine the relationship between gender and collaboration behaviour, we make an adjustment to the collaboration statistics. Specifically, we use model residuals to capture the collaboration behaviour with the effects of the publication number removed (further details given in supplementary material~\ref{adjusted_collaboration})~\cite{dormann2013collinearity}. We denote these statistics that account for the potential influence of publication count as ``adjusted''.
\begin{table}[H] 
\vspace{-7pt}
\centering 
\caption{\textbf{Regression coefficients indicate a significant relationship between the number of papers and collaborative behaviour statistics.} Each column represents a distinct regression model, with the type of regression (normal or logistic) highlighted for each dependent variable. The values in parentheses below each coefficient denote the standard errors.} 

\scalebox{0.75}{
\begin{tabular}{@{\extracolsep{5pt}}l c c c c} 
\\[-1.8ex]\hline 
\hline \\[-1.8ex] 
 & \multicolumn{4}{c}{\textit{Dependent variable:}} \\ 
\cline{2-5} 
\\[-1.8ex] & \multicolumn{1}{c}{network size} & \multicolumn{1}{c}{tie strength} & \multicolumn{1}{c}{clustering coefficient} & \multicolumn{1}{c}{co-authors per publication} \\ 
\\[-1.8ex] & \multicolumn{1}{c}{\textit{normal}} & \multicolumn{1}{c}{\textit{normal}} & \multicolumn{1}{c}{\textit{logistic}} & \multicolumn{1}{c}{\textit{normal}} \\ 
\\[-1.8ex] & \multicolumn{1}{c}{(1)} & \multicolumn{1}{c}{(2)} & \multicolumn{1}{c}{(3)} & \multicolumn{1}{c}{(4)}\\ 
\hline \\[-1.8ex] 
no.\ of\ papers & 0.036$^{***}$ & 0.007$^{***}$ & $-$0.07$^{***}$ & $-$0.007$^{***}$ \\ 
 & (0.0003) & (0.0002) & (0.003) & (0.001) \\ 
Constant & 2.125$^{***}$ & 0.459$^{***}$ & $-$0.197$^{***}$ & 2.918$^{***}$ \\ 
 & (0.007) & (0.003) & (0.029) & (0.014) \\ 
\hline \\[-1.8ex] 
\textit{Note:} & \multicolumn{4}{r}{$^{*}$p$<$0.1; $^{**}$p$<$0.05; $^{***}$p$<$0.01} \\ 
\end{tabular}}
\label{regression_papers}
\vspace{-7pt}
\end{table} 

\subsection{Correlation between collaboration statistics}
\begin{figure}[H]
\vspace{-7pt}
\centering
\includegraphics[width=0.9\textwidth]{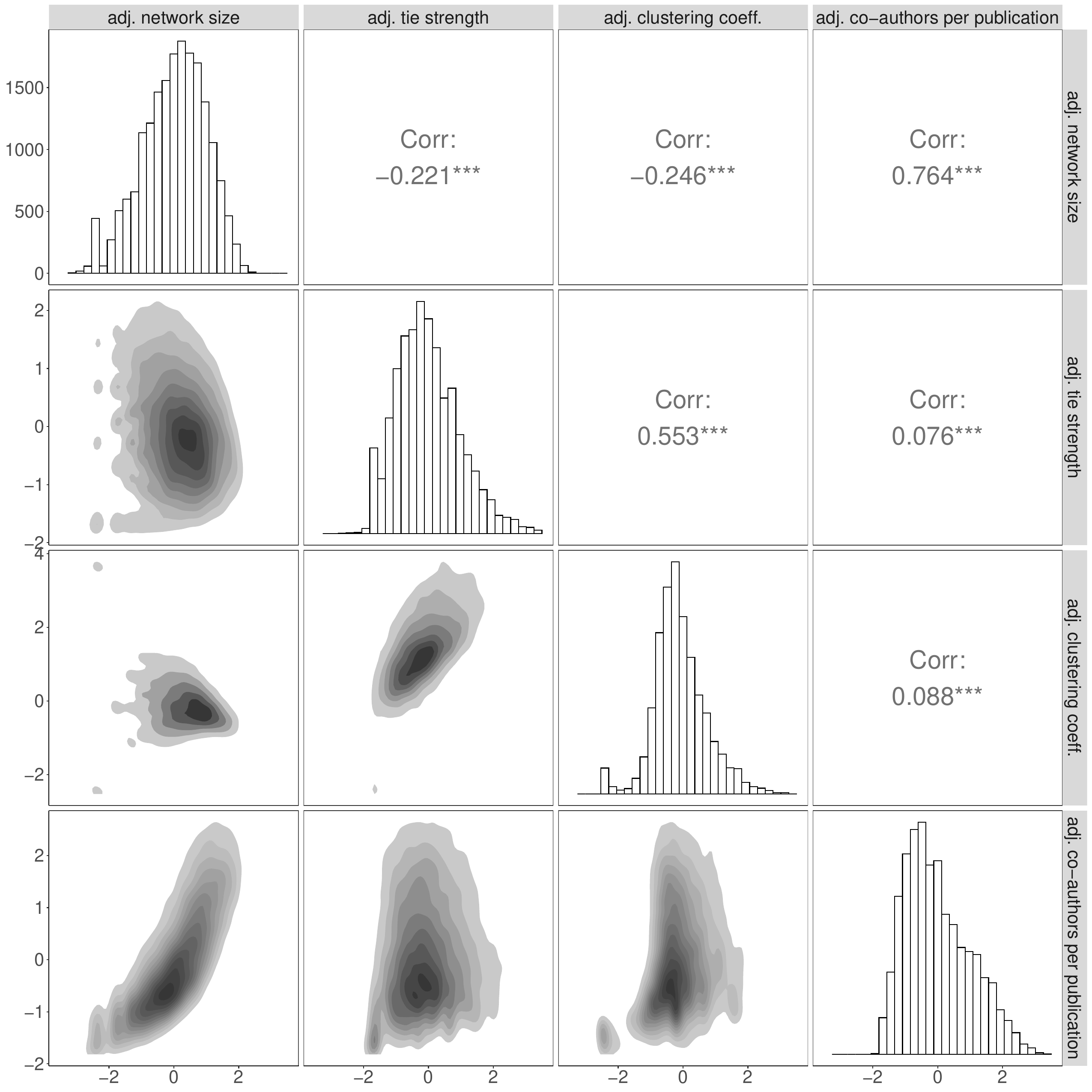}  
\caption{\textbf{Distribution and correlation of adjusted collaboration statistics}. On the diagonal, histograms show the distribution for each statistic. The lower left shows density heat maps of the joint distributions of pairs of statistics. Darker shades indicate higher density. The upper right shows the corresponding correlation coefficients. The presence of asterisks (***) indicates statistical significance of the correlation coefficients ($p < 0.05$). }
\label{corre_plot}
\vspace{-7pt}
\end{figure}

Here we examine the correlation between the different collaboration statistics to better understand how they relate to each other. Figure~\ref{corre_plot} displays the pairwise correlations between adjusted collaboration statistics. 

All pairs of collaboration statistics are found to be statistically significant, however, for most pairs, the effect size is relatively small as the estimated coefficients are close to zero. Exceptions are network size with co-authors per publication and tie strength with clustering coefficient. Since these statistics are adjusted for the number of publications, which remove the effect of publication count, it is expected that the network size would correlate with the mean number of co-authors per publication. When an author has a larger network size, they tend to have more co-authors per paper. 
The moderate correlation between tie strength and clustering coefficient seems to indicate that repeated co-authorship is associated with triadic closure, i.e., co-authors become more interconnected.
We also see that tie strength and clustering coefficient have a weak negative correlation with network size. This relationship is likely due to the fact that as the network size increases, the collaboration network can become less dense and repeated collaborations occur less frequently, after adjusting for the number of publications.
\subsection{Gender differences in collaboration behaviour}
In order to investigate the difference in collaborative behaviour between male and female authors, we examine both the distributions of collaborative behaviour and the mean collaboration statistics derived from Generalised Linear Models (GLMs). This dual approach helps us to discern potential disparities in collaboration patterns between male and female authors.

\begin{figure}[H]
\vspace{-7pt}
\centering
\includegraphics[width=1\textwidth]{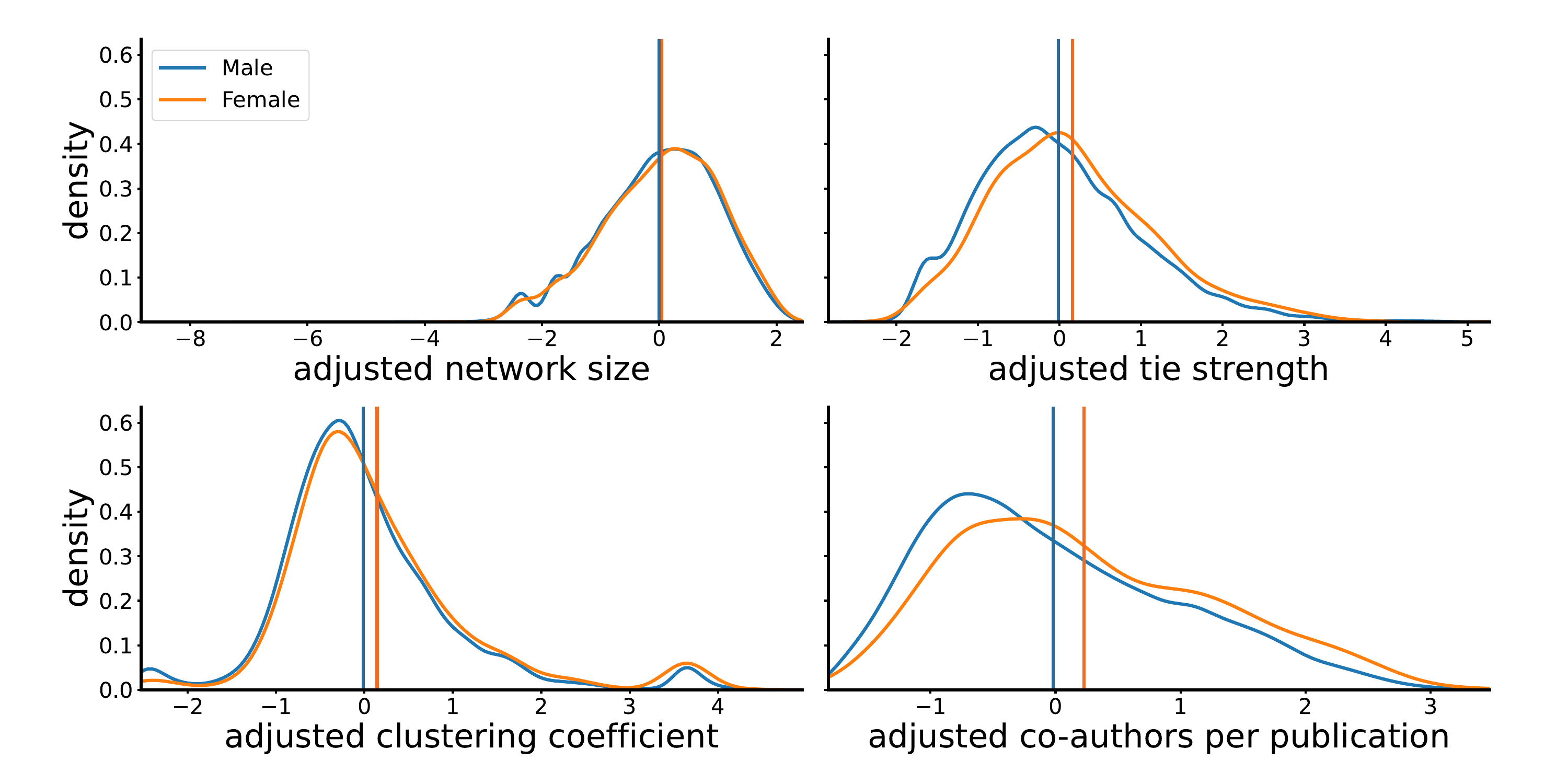}  
\caption{\textbf{Gender differences in collaboration statistics.} We show the distribution of collaboration statistics. Blue curves represent male scientists and orange curves represent female scientists. Vertical lines indicate the mean value. }
\label{collaboration_differences}
\vspace{-7pt}
\end{figure}

Figure~\ref{collaboration_differences} presents the density distributions of adjusted collaborative behaviour statistics for both genders, normalised based on the number of publications.
These visualisations reveal that the distributions of network size are indistinguishable across genders when accounting for the number of published papers.
On the contrary, female authors seem to have greater mean tie strength, clustering coefficient and average number of co-authors per publication compared to male authors.
GLM coefficients (see supplementary~\ref{GLM_adjusted} for details) confirm these differences in means to be statistically significant. 

\subsection{Assortative mixing patterns in collaborations}
Here we investigate to what extent collaboration behaviour is similar between collaborators. For instance, focal authors who frequently collaborate together might also start to collaborate in similar ways. 
We use network assortativity~\cite{newman2003mixing} to assess the extent to which focal authors tend to exhibit similar collaboration characteristics as their co-authors.
We stratify the author pairs into distinct collaboration tiers based on the number of papers they co-authored. At each tier, we calculate the assortativity of the collaboration statistics between co-authors to determine whether or not the frequency of collaboration affects the level of assortativity.
Note that assortativity values are based on the residual collaboration statistics once we remove the author pair under consideration. 
For example, when calculating the assortativity of mean tie strength between a pair of authors $a$ and $b$, we calculate their mean tie strengths without including the tie between $a$ and $b$.

\begin{figure}[H]
\vspace{-7pt}
\centering
\includegraphics[width=1\textwidth]{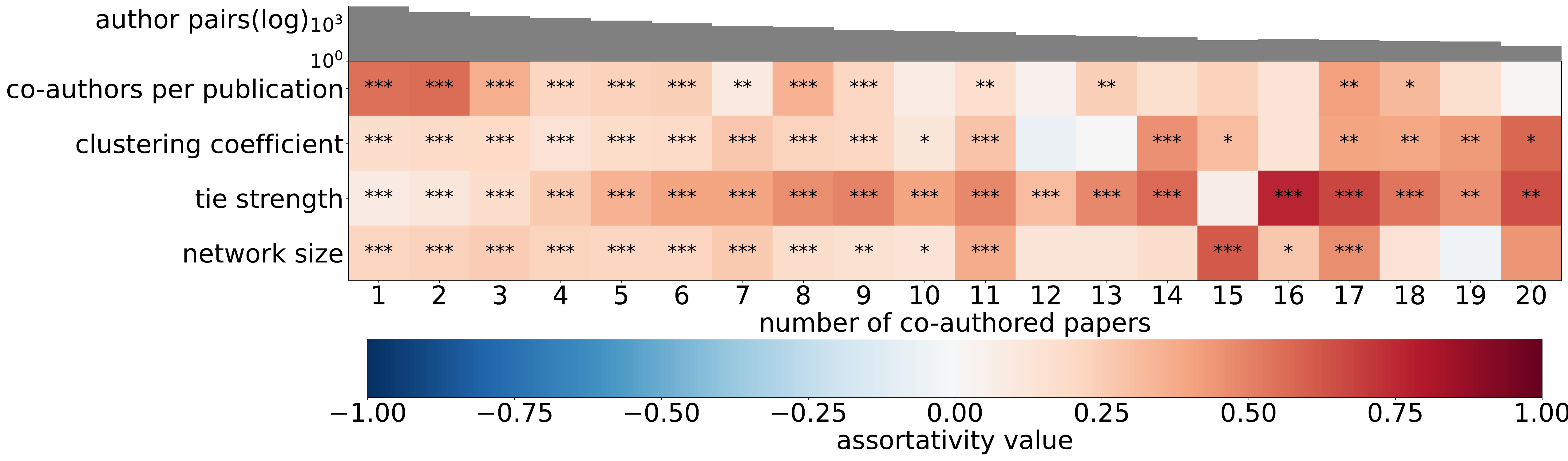}  
\caption{\textbf{Assortativity of collaboration statistics according to the number of co-authored publications.} The heat map provides a visual representation of the assortativity of each of the collaboration statistics (y-axis) between co-author pairs.
The bar chart along the top shows the logarithmic frequency of the co-authorship pairs. A positive assortativity value indicates that individuals with similar collaboration behaviour are more likely to collaborate with each other.}
\label{assort_colla}
 \vspace{-7pt}
\end{figure}

Figure~\ref{assort_colla} shows the generally positive assortativity for all four collaboration statistics, indicating that authors with similar collaboration styles tend to collaborate together.
The tie-strength assortativity grows with the number of co-authored publications.
Often repeated ties link pairs of scientists with similar average tie strengths in their local networks.
In contrast, the assortativity of co-authors per publication decreases with repeated collaboration.
Co-authors that publish only once or twice are most similar in terms of their average number of co-authors per publication.

\section{Academic career progression}
To measure academic career progression, we calculate three statistics that capture the duration of a focal author's career and the time spent at different career stages, i.e., before/after becoming a principal investigator (PI; see also Figure~\ref{fig:fig1}). 
\begin{itemize}
    \item \textbf{career length}: the number of years between a focal author's first and final publication.
    \item \textbf{time to become a PI}: the number of years between an author's first publication and their third last-author publication. 
    \item \textbf{career length after becoming a PI}: the number of years between a focal author's third last-author publication and their final publication.
\end{itemize}

Note that we use the third last-author publication to estimate the time when a focal author transitions to being a PI. In physics, author order is often used to indicate relative seniority where more junior contributors are listed earlier and more senior collaborators, i.e., those in supervisory roles, appear later in the author list. We therefore expect that an author's transition from first to last author is an indication of their seniority. However, this convention is not universal, as some authors opt for alphabetical ordering. These authors may appear to have reached PI status simply because they appear last by alphabetical order.
We therefore exclude papers, when estimating the transition to PI, in which the authors are alphabetically ordered (see supplementary material~\ref{identify_PI} for more information). 

\subsection{Gender differences in career progression}
To investigate gender differences in academic career progression, we examine three key aspects: the likelihood of becoming a PI, overall career length and career length after becoming a PI. We use the cumulative distribution function to show the probability of becoming a PI for each gender by a given time. 
We visualise the overall career length and career length after PI using the complementary cumulative distribution function for male and female authors. 
We can assess whether there are profound disparities in career progression between male and female authors by comparing these curves.

\begin{figure}[H]
\vspace{-7pt}
\centering
\includegraphics[width=1\textwidth]{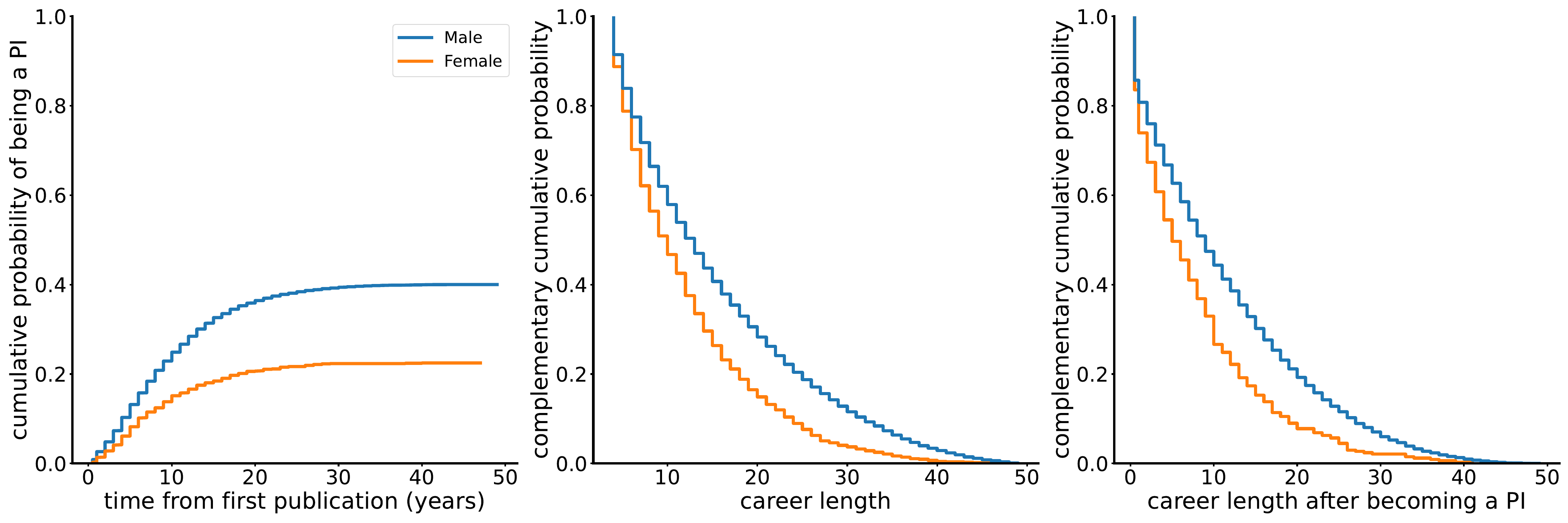}  
\caption{\textbf{Gender differences in career progression.} 
These graphs show cumulative probability of being a PI and complementary cumulative probability during career length and career length after PI. Solid coloured lines represent cumulative and complementary cumulative density functions.}
\label{KM_curve}
\vspace{-7pt}
\end{figure}

Figure~\ref{KM_curve} shows that female authors have a lower probability of becoming a PI compared with male authors. Male and female authors also have a clear gap in career length and career length after PI, with the male authors displaying greater longevity in both cases. This shows that female authors have had a lower probability of sustained career length, both in terms of overall career duration and after achieving PI status.

\subsection{Assortativity in career progression}
We also explore the assortativity of career progression between focal authors, according to the number of co-authored publications. In other words, do authors who regularly collaborate with other scholars on publications exhibit similar trajectories in terms of career progression? 

\begin{figure}[H]
\vspace{-7pt}
\centering
\includegraphics[width=1\textwidth]{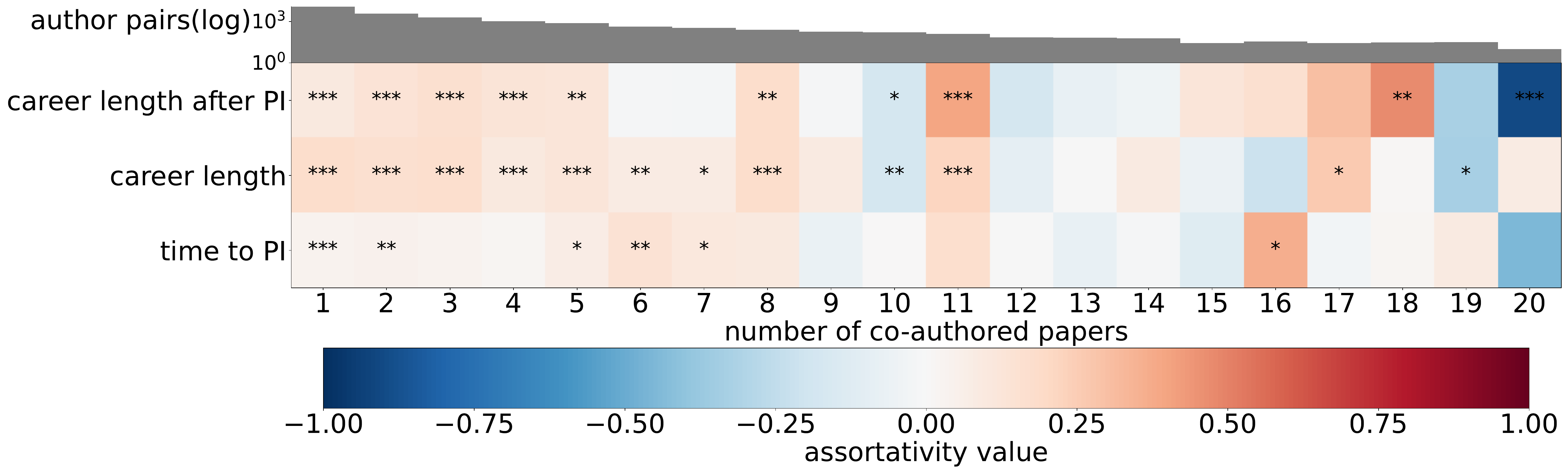}  
\caption{\textbf{Assortativity of career progression statistics according to the number of co-authored publications.} The heat map provides a visual representation of the assortativity of each of the career progression statistics (y-axis) pairs of co-authors. The colour of each cell represents the assortativity according to the number of co-authored publications (x-axis). The bar chart along the top shows the logarithmic frequency of the co-authorship pairs. A positive assortativity value indicates that individuals with similar career progression are more likely to collaborate with each other.}
\label{assort_career}
\vspace{-7pt}
\end{figure}

Figure~\ref{assort_career} shows the assortativity of time spent at different career stages between co-authors. Overall we see slight positive assortativity in the career progression statistics. There does not appear to be any clear pattern in relation to the number of co-authored publications. These results suggest that frequent collaboration has little or no effect on the similarity of a co-author's career progression. To validate the observed patterns in assortativity, we conducted a complementary analysis based on pairwise similarity in career outcomes between co-authors. As detailed in Appendix~\ref{pairwise_similarity_section}, the results confirm that higher collaboration frequency does not necessarily imply similar career progression.

\section{Relationship between collaboration behaviour and career progression}
Now we explore the relationship between collaboration behaviour, career progression and whether or not this relationship varies by gender. We use logistic regression models to investigate the relationship between collaboration statistics and the likelihood of an individual becoming a Principal Investigator (PI) and Accelerated Failure Time Models (AFTs) to analyse the association between collaboration and the time spent at different career stages.

\subsection{Likelihood of becoming a PI}
We assess the relationship between each of the adjusted collaboration statistics (network size, tie strength, clustering coefficient and co-authors per publication) and the likelihood of becoming a PI using logistic regression models. We also include gender and the interaction between gender and the collaboration statistic as predictors. We model the probability of an individual $i$ succeeding to become a PI ($s_i=1$) as follows:
\begin{equation}
\textrm{P}(s_i=1 \mid g_i, c_i) = \frac{1}{1+e^{(-\beta_0 - \beta_1 c_i - \beta_2 g_i - \beta_3 g_ic_i)}} \enspace , \label{eq_logreg}
\end{equation}
where $g_i$ is a binary variable that denotes the gender of the $i$-th individual, where $g_i = 1$ indicates female and $g_i = 0$ indicates male. The variable $c_i$ corresponds to the value of the specific collaboration statistic for the $i$-th individual. The model includes $\beta_0$ as the intercept term, $\beta_1$ for the collaboration static effect, $\beta_2$ for the gender effect and $\beta_3$ for the interaction effect between gender and the collaboration statistic. We also construct two separate models, one for male and one for female focal authors, to analyse the relationship within the two gender groups by fixing $\beta_2 = \beta_3 = 0$.

\begin{figure}[H]
\vspace{-7pt}
\centering
\includegraphics[width=0.99\textwidth]{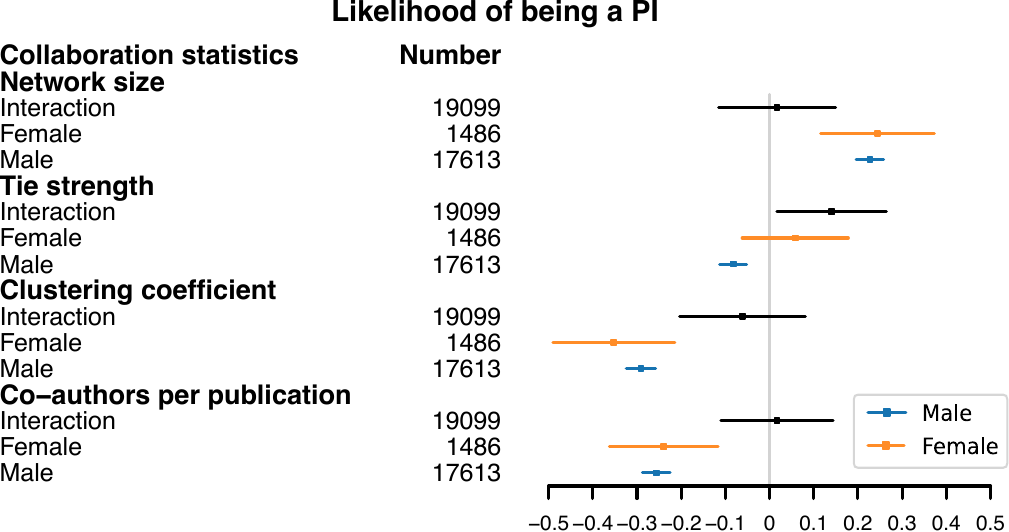}  
\caption{\textbf{The forest plot of adjusted collaboration statistics on likelihood of being a PI.} We present three types of model for each type of collaboration statistic (network size, tie strength, clustering coefficient and co-authors per publication). The \textit{interaction} model refers to the full model described in Eq.~\eqref{eq_logreg} and includes the interaction effect of gender with the collaboration statistic. The \textit{female} and \textit{male} models assess the effect of the respective collaboration statistic on the likelihood of being a principal investigator (PI) for the separate groups of female and male authors. The \textit{Number} column shows the sample size of each model. The markers show the coefficient estimates, with error bars indicating the 95\% confidence intervals (CIs) for each estimate.}
\label{Likelihood}
\vspace{-7pt}
\end{figure}
Figure~\ref{Likelihood} is a forest plot illustrating the association between the adjusted collaboration statistics and the likelihood of becoming a Principal Investigator (PI). The plot includes three types of model (\textit{interaction}, \textit{female} and \textit{male}) for each adjusted collaboration statistic (network size, tie strength, clustering coefficient and co-authors per publication). Each row represents the value of the relevant model coefficient given in Eq.~\eqref{eq_logreg}. 
In the \textit{interaction} models, we include the coefficient $\beta_3$ to test whether the effect of each collaboration statistic on the likelihood of becoming a PI differs between male and female authors. A non-significant interaction term ($\beta_3$) indicates that the effect does not significantly vary across genders. In this specification, $\beta_1$ represents the effect of the collaboration statistic for male authors (the reference group), while the effect for female authors is given by the sum $\beta_1 + \beta_3$, as also shown in the separate \textit{male} and \textit{female} models.

Figure~\ref{Likelihood} shows that the interaction terms between gender and most collaboration statistics are not statistically significant, except for tie strength. For men, a weaker mean tie strength is significantly associated with a greater likelihood of becoming a PI. However, the smaller number of female authors makes it harder to assess the statistical significance between mean tie strength and the likelihood of being a PI for women.
We also see that an increased network size is significantly associated with a higher likelihood of becoming a PI, while a lower clustering coefficient and number of co-authors per publication are significantly associated with a greater likelihood of becoming a PI.

\subsection{Duration of career stages}
We examine the impact of collaborative behaviour and gender on career progression using accelerated failure time (AFT) models. Specifically, we implement three sets of AFT models, each set predicts the duration of different career stages: time to attain PI status, overall career length and career length after becoming a PI. In each set, we fit an AFT model for each collaborative behaviour statistic, for both genders and the interaction between genders, just as we did when estimating the likelihood of becoming a PI in the previous section. The AFT model is written as follows:
\begin{equation}
\log(T_i \mid g_i, c_i) = \beta_0 + \beta_1 c_i + \beta_2 g_i + \beta_3 (g_ic_i) + \epsilon_i  \enspace , \label{eq_surreg}
\end{equation}
where $T_i$ denotes the time for a focal author to become a PI, their career length, or their career length after becoming a PI. The model includes $\beta_0$ as the intercept term, $\beta_1$ for the collaboration static effect, $\beta_2$ for the gender effect and $\beta_3$ for the interaction effect between gender and the collaboration statistic. Finally, $\epsilon_i$ is an error term.

\subsubsection{Overall career length}
\begin{figure}[H]
\vspace{-7pt}
\centering
\includegraphics[width=1\textwidth]{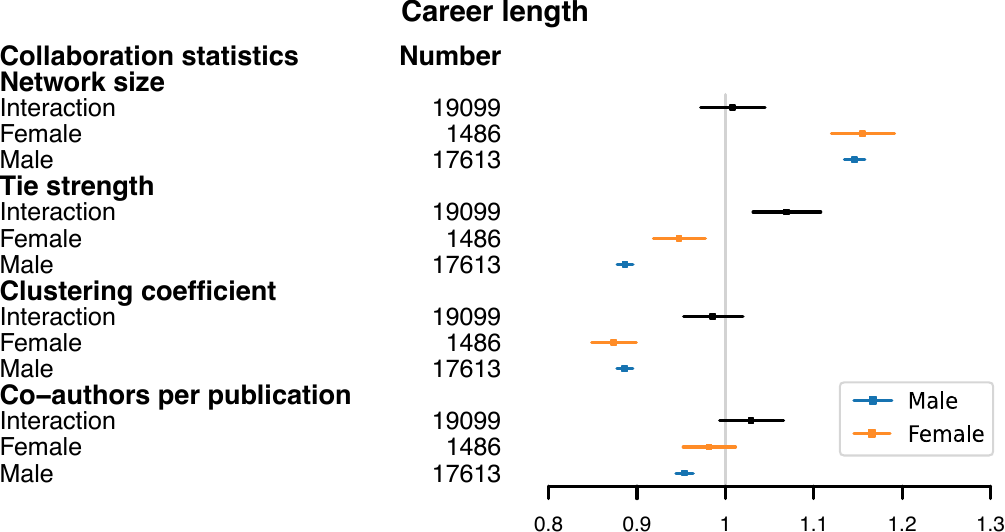}  
\caption{\textbf{The forest plot of adjusted collaboration statistics on career length.} We present three types of model for each type of collaboration statistic (network size, tie strength, clustering coefficient and co-authors per publication). 
The \textit{interaction} model refers to the full model described in Eq.~\eqref{eq_surreg} and includes the interaction effect of gender with the collaboration statistic. The \textit{female} and \textit{male} models assess the effect of the respective collaboration statistic on career length for the separate groups of female and male authors. The \textit{Number} column shows the sample size of each model. The markers show the coefficient estimates, with error bars indicating the 95\% confidence intervals (CIs) for each estimate.}
\label{career_length}
\vspace{-7pt}
\end{figure}

Figure~\ref{career_length} shows the interaction between gender and adjusted tie strength is found to be significant, suggesting that a weaker mean tie strength is associated with longer careers, but this relationship is less pronounced for females compared to males. 
We see that overall having a larger network size appear to be associated with a longer academic career, while lower clustering coefficient and fewer co-authors per publication seem to be linked to a longer career length. These findings appear to be consistent between genders and indicate that a wider collaboration network, with weaker ties and clustering might contribute to career longevity. 

\subsubsection{Time to become a principal investigator and career length after PI}
\begin{figure}[H]
\vspace{-7pt}
\centering
\includegraphics[width=1\textwidth]{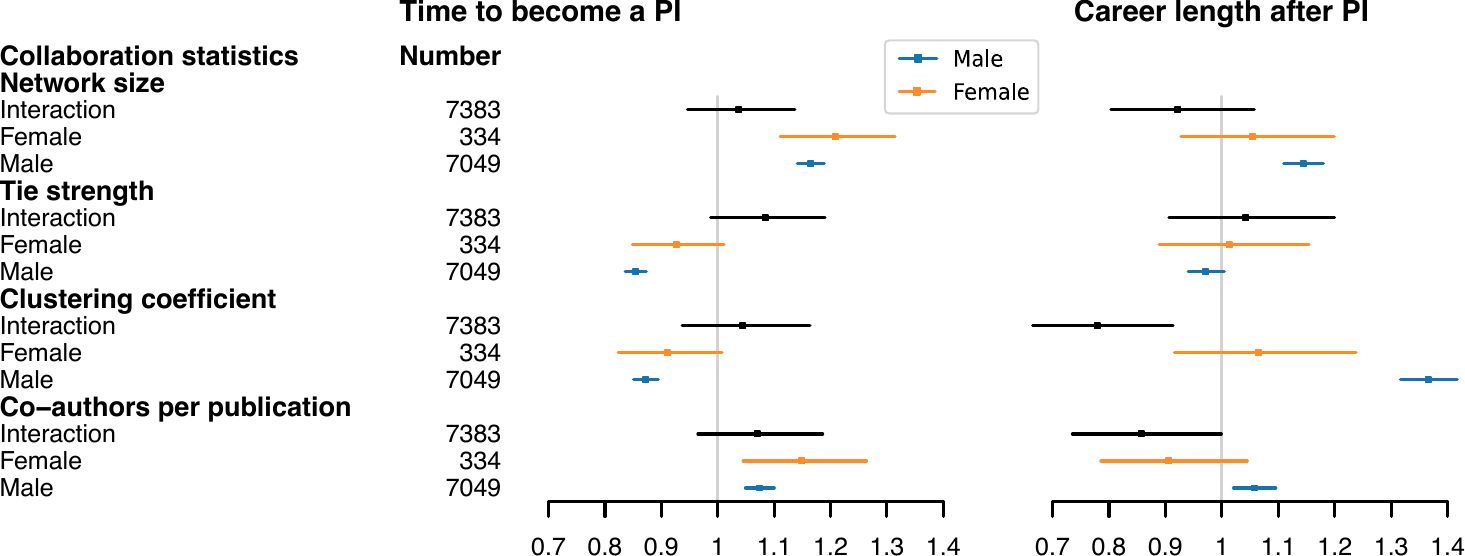}  
\caption{\textbf{The forest plots of adjusted collaboration statistics on Time to become a PI and Career length after PI.} 
In each forest plot, we present three types of model for each type of collaboration statistic (network size, tie strength, clustering coefficient and co-authors per publication).
The \textit{interaction} model refers to the full model described in Eq.~\eqref{eq_surreg} and includes the interaction effect of gender with the collaboration statistic. The \textit{female} and \textit{male} models assess the effect of the respective collaboration statistic on time to become a PI/career length after PI for the separate groups of female and male authors. The \textit{Number} column shows the sample size of each model. The markers show the coefficient estimates, with error bars indicating the 95\% confidence intervals (CIs) for each estimate.}
\label{timepi_clpi}
\vspace{-7pt}
\end{figure}

Figure~\ref{timepi_clpi} shows the interaction terms between gender and the four collaboration statistics are not statistically significant except between gender and clustering coefficient in career length after PI.
The adjusted collaboration statistics have a significant relationship with the time to becoming a PI (Figure~\ref{timepi_clpi}). Specifically, we see that an increase in network size and number of co-authors per publication tends to prolong the expected time to becoming a PI. On the contrary, weaker mean tie strength and clustering coefficient seem to be associated with prolonged time to becoming a PI for male authors.

In Figure~\ref{timepi_clpi} we see that a larger network size, a higher clustering coefficient and a greater number of co-authors per publication are positively correlated with career length post-PI, but this relationship is more ambiguous among the female authors due to the small sample size. However, our additional analyses (given in supplementary material~\ref{collaboration_career_PI}) suggest that having a larger network size, higher clustering coefficient and greater average number of co-authors at the time of becoming a PI is negatively associated with career length post-PI. This suggests that tightly-knit or larger-scale collaborations at the time of becoming PI might limit opportunities for sustained academic development, even though final higher clustering may contribute to extending a career in some contexts.
In addition to these collaboration statistics, we also explored the gender composition of ego networks as a supplementary collaboration statistic (see more details in Appendix ~\ref{female_ratio}). We find that female authors tend to collaborate with a higher proportion of female co-authors and that female ratio in co-authorship networks shows a weak association with career outcomes.

\section{Discussion}
Here we have examined the collaboration behaviour and career progression of female and male focal authors in physics.  
Our findings indicate that the relationship between collaboration behaviour and career progression is largely the same for both women and men. 
However, despite this relationship being consistent across genders, we do observe differences in the way men and women collaborate and how they progress in their careers. 
Women are more likely than men to publish repeatedly with the same co-authors and to participate in larger, more interconnected collaborations, a pattern also observed in economics~\cite{ductor2023gender}.
These collaborative behaviours, regardless of gender, are linked to negative career progression outcomes. However, women exhibit these less favourable collaborative patterns more frequently and also have poorer career outcomes.

We find that those who eventually reach the seniority of principal investigator (PI), tend to have published with more unique collaborators. In contrast, publishing repeatedly with the same highly interconnected collaborators and/or larger number of co-authors per publication is characteristic of shorter career lengths and those that do not attain PI status. These findings align well with the suggestions that prolonged engagement in specialised or internally-focused collaborations can become harmful~\cite{pike2010collaboration} and that publishing with a large number of co-authors per publication may dilute individual contributions and visibility~\cite{ioannidis2008measuring}. However, at the same time, we also see that those that publish repeatedly with the same set of interconnected collaborators and do attain the status of PI, do so in less time. This is in agreement with recent findings on the benefits of highly repetitive collaborations for career progression~\cite{petersen2015quantifying}.

Our findings align with prior research, such as the study by Van der Wal et al.~\cite{Jessica2021collaboration}, in highlighting the importance of collaboration behaviour in relation to career progression. Lindenlaub and Prummer~\cite{lindenlaub2021network} demonstrate that tighter networks hinder productivity by restricting information flow and novel idea exchange, factors critical for career progression. 
Both studies observe that larger and more diverse networks are beneficial for career advancement, while tightly clustered networks (indicated by higher clustering coefficients) may limit diversity in collaborations and visibility within broader professional communities. 
Our observation that smaller, tightly interconnected networks are associated with reduced likelihood of transitioning to PI status is consistent with these studies. Moreover, this pattern remains consistent across subfields in physics (see~\ref{Journal_based_analysis} for further details).
However, we also identify that more clustered networks have an accelerated transition to PI status in physics. 
In contrast, the study by Van der Wal et al. associates less clustered networks with faster PI transitions in behavioural ecology. This difference may reflect variations in the collaborative and publishing norms of these disciplines. Furthermore, while Van der Wal et al. identify stronger positive relationships between network statistics and career progression for women, our results do not indicate a significant difference between genders.

Close-knit collaboration networks may be good for fostering professional support~\cite{brown1987social, krackhardt2003strength} and rapid project completion~\cite{de2014strength}. However, they may also reduce opportunities to form new, potentially more influential, professional connections~\cite{skilton2010effects, barabasi2002evolution, michelfelder2013and} and reduce visibility and recognition within the broader scientific community~\cite{lee2019homophily}, thereby impacting career advancement.
Furthermore, the preference for larger and more interconnected groups may also dilute individual contributions, making it harder for female researchers to stand out as leaders in their field.
Higher dropout rate of women is closely related to the well-documented ``leaky pipeline'' which happens at all levels of the profession, from undergraduates to faculty~\cite{figueiredo2023walking, martinez2021women, fry2021stem, mehta2022faculty, stoet2018gender, de2019status}, hampering career progression for women. The lack of career progression may then influence the opportunities for other types of collaborative behaviour.

Our results can potentially enhance existing policies that address gender balance in scientific collaboration. For instance, Horizon Europe, a major funding instrument in the European Union requests details regarding the ``gender dimension'' of all grant proposals~\cite{horizon_europe_2023}.
First, policies could be implemented to facilitate opportunities for female researchers to engage with a wider array of collaborators outside their immediate circle, to align with our finding that a greater number of collaborators is positively associated with career progression. One major obstacle is the lower mobility of women compared to men, which substantially limits their network size~\cite{momeni2022many}. Lowering restrictions on mobility might be achieved by providing hybrid working conditions, flexible multi-stage childcare support, or financial incentives that encourage mobility. 
Second, policies that reward impactful small-team publications could significantly enhance the visibility and recognition of female researchers' contributions. These measures could include adjustments in performance evaluations to highlight individual work, dedicated funding for projects led by women, mentorship programs that support small teams and workshops or training programs focused on developing skills for leading smaller research teams.  
Third, recognising and rewarding diverse collaboration efforts in performance evaluations and grant applications can motivate researchers to seek out varied and expansive professional relationships, ultimately supporting their career advancement.

Here we have focused on statistics of the careers of authors in physics based only on publication data from the American Physical Society and therefore do not have a complete publication record for all these focal authors. While this necessarily means that we do not have the complete collaboration network for each author, it seems unlikely that collaboration behaviour would be significantly different when publishing in other journals. 
Additionally, inferring gender from names and principal investigator (PI) status from being the last author may be inaccurate due to factors such as non-binary names, cultural differences in naming conventions and variations in authorship practices across fields and institutions. However, these methods are comparable to the state of the art and have been used in other studies~\cite{liu2013s, karimi2016inferring, sebo2021using, Jessica2021collaboration}. 
We find various associations between collaborative behaviours and career progression outcomes, however we cannot currently determine the direction of causality or even if a causal relation exists.
Moreover, an author's collaborative behaviour may not be consistent throughout different times of their career, which might not capture the intricacies and shifts in collaboration patterns across different career stages. For instance, an author might be highly collaborative during their early career and work more isolated later on, or vice versa. Such nuances get lost when looking at career-span averages. Future research could benefit from segmenting the career into different phases and analysing collaboration patterns within each phase to provide a more nuanced understanding. 

\section*{Acknowledgment}
The authors thank Jessica E. M. van der Wal for helpful conversations. MS acknowledges the financial support from the China Scholarship Council under grant number 202108080241. JB was supported by the Austrian Science Promotion Agency FFG under project No. 873927 ESSENCSE.
JB is a recipient of a DOC Fellowship of the Austrian Academy of Sciences at the Complexity Science Hub. LP was supported in part by the Dutch Research Council (NWO) Talent Programme ENW-Vidi 2021 under grant number VI.Vidi.213.163.

\section*{Data availability}
The APS datasets are available upon request to the American Physical Society (\url{https://journals.aps.org/datasets}).
The data analysed in this study have been deposited in the Dryad Digital Repository and can be accessed via \url{https://doi.org/10.5061/dryad.b5mkkwhr4}.
Relevant code for this research work are stored in GitHub: \url{https://github.com/mingrongshe/Gender_differences_Physics}
and have been archived within the Zenodo repository: 
\url{https://doi.org/10.5281/zenodo.13739434}.
Before publication, the code will be further developed to generate and integrate these intermediate results.



\bibliography{main}

\clearpage 
\cleardoublepage
\appendix

\setcounter{figure}{0}
\renewcommand{\thefigure}{S\arabic{figure}}
\setcounter{section}{0}
\renewcommand{\thesection}{S\arabic{section}}
\setcounter{table}{0}
\renewcommand{\thetable}{S\arabic{table}}
\renewcommand{\theHtable}{S\arabic{table}}
\renewcommand{\theHfigure}{S\arabic{figure}}
\renewcommand{\theHsection}{S\arabic{section}}

\section*{Supplementary Material}
\section{Data preprocessing}
\label{data_process}
\subsection{Data collecting}
We obtain the dataset from the American Physical Society (APS) between 1893 to 2020. The dataset is comprised of all the publications in Physical Review, including 19 different journals: PRA, PRAB, PRAPPLIED, PRB, PRC, PRD, PRE, PRFLUIDS, PRI, PRL, PRMATERIALS, PRPER, PRRESERACH, PRSTAB, PRSTPER, PRX, PRXQUANTUM and RMP. 

\begin{table}[H] 
\centering
\caption{Summary of journals published by the American Physical Society (APS) from the years 1893 to 2020. The table shows the subject focus of each journal, its respective impact factor (IF) as per the 2022 Journal Citation Reports (Clarivate Analytics, 2022) and the total number of papers published in each journal up to 2020. The acronyms of the journals are provided alongside their full names for easy reference.}
\resizebox{\linewidth}{!}{
\begin{tabular}{l l l c c}
\toprule 
Acronym & Name & Subject & IF & Papers \\
\midrule 
PR & Physical Review &  Aspects of Physics & - & 47,940 \\
PRA & Physical Review A & Atomic, Molecular and Optical Physics & 2.9 & 83,632 \\
PRAB & Physical Review Accelerators and Beams & Accelerator science, Technology and Applications & 1.7 & 1,175 \\
PRAPPLIED & Physical Review Applied & Applied Physics & 4.6 & 3,536 \\
PRB & Physical Review B & Condensed Matter and Materials Physics & 3.7 & 197,564 \\
PRC & Physical Review C & Nuclear Physics & 3.1 & 41,895 \\
PRD & Physical Review D & Particles, Fields, Gravitation and Cosmology & 5.0 & 94,833 \\
PRE & Physical Review E & Statistical, Nonlinear, Biological and Soft Matter Physics & 2.4 & 61,903 \\
PRFLUIDS & Physical Review Fluids & Fluid Dynamics & 2.7 & 2,375 \\
PRI & Physical Review I & (Not a regular journal) & - & 1,469 \\
PRL & Physical Review Letters & Cross-disciplinary Physics Research & 8.6 & 129,116 \\
PRMATERIALS & Physical Review Materials & Materials Research & 3.4 & 2,532 \\
PRPER & Physical Review Physics Education Research & Physics Education Research & 3.1 & 481 \\
PRRESEARCH & Physical Review Research & Cross-disciplinary Physics Research & 4.2 & 2,380 \\
PRSTAB & Physical Review Special Topics - Accelerators and Beams & Special Topics in Accelerator and Beam Physics & 1.6 & 2,399 \\
PRSTPER & Physical Review Special Topics - Physics Education Research & Special Topics in Physics Education Research & 2.4 & 368 \\
PRX & Physical Review X & Cross-disciplinary Physics Research & 12.5 & 1,857 \\
PRXQUANTUM & Physical Review X Quantum & Quantum Information Science & 9.7 & 38 \\
RMP & Reviews of Modern Physics & Review articles on Physics & 44.1 & 3,423 \\
\bottomrule 
\end{tabular}}
\label{APS_data}
\end{table}

\subsection{Name disambiguation}
Name disambiguation is an important and often challenging first step in analysing scientists' careers. The difficulty arises from the fact that multiple individuals can share the same name, while a single individual may appear under different name variants.
The goal of name disambiguation is to accurately establish authorship links between authors and their respective publications using the available metadata. Here, we use an author's name, affiliation and their collaborators to determine if two publications belong to the same or to different identities. We use a process based on prior research by Sinatra et al.~who applied such an algorithm to the APS dataset~\cite{sinatra2016quantifying}. The data is first filtered to exclude papers with more than $10$ authors and those without at least one affiliation per author in order to ensure sufficient contribution by each author.
Subsequently, all author names are grouped if they have identical surnames and share matching initials. Sinatra et al. then merge the author identities of such author pairs if they meet one of the following criteria: (1) the authors cited each other at least once; (2) they shared at least one co-author; (3) they shared at least one affiliation.
To avoid ambiguities in the definition of the merging step, we follow a recently proposed implementation of this algorithm~\cite{bachmann2024cumulativeadvantagebrokerageacademia}.
This solution follows the scheme provided by Sinatra et al. and merges matching author pairs only if their first names share the first initial and they are not conflicting with previously assigned first names.

\subsection{Gender inference}
Since the APS dataset does not include author gender, we employ a recently developed method to infer genders based on authors' names~\cite{buskirk.etal_opensourceculturalconsensus_2023}.
This method performs on-par with alternative solutions, but has the advantage that it follows open data and open source paradigms.
That is, both source code and data are openly available.
In addition, it was recently applied in a comparable study~\cite{bachmann2024cumulativeadvantagebrokerageacademia}.
In an iterative procedure, the inference method assigns trust levels for their sources of data based on their agreement on gender assignments with the other sources.
These trust scores then weight the decision on the final gender assignment of a given name.
Repeating this procedure until both the gender inference decisions and the trust scores converge yields an estimate of the trustworthiness of a source, as well as a probabilistic gender assignment.
We assign gender to author names when a certainty of at least $66\%$ can be guaranteed.
Finally, we aggregate the assignment for multiple author names based on a majority vote.
In case of a draw between male and female assignment, or when no gender could be inferred for any name, we assign an `unknown' gender label. By following this rule, we identify the gender of $248,788$ authors.

\begin{figure}[H]
\centering
\includegraphics[width=0.99\textwidth]{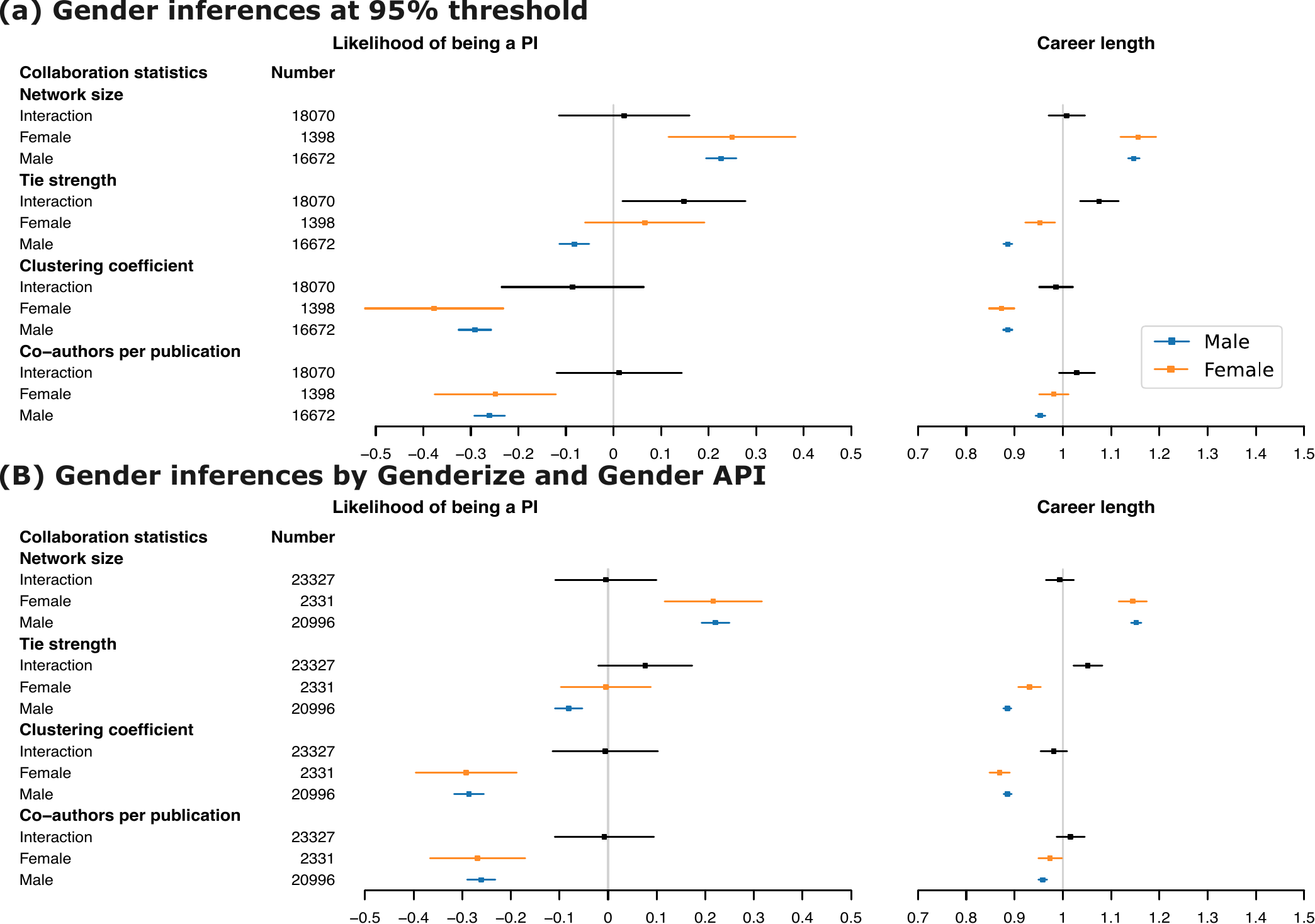}  
\caption{\textbf{The forest plots of adjusted collaboration statistics on likelihood of being a PI and career length using different gender inference methods.} The plots compare results obtained under two gender inference methods. (a) shows the analysis with a stricter 95\% probability threshold for gender assignment, while (b) shows results derived from Genderize and Gender API. 
In each forest plot, we present three types of model for each type of collaboration statistic (network size, tie strength, clustering coefficient and co-authors per publication).
The \textit{interaction} model refers to the interaction effect of gender with the collaboration statistic. The \textit{female} and \textit{male} models assess the effect of the respective collaboration statistic on likelihood of being a PI/career length for the separate groups of female and male authors. The \textit{Number} column shows the sample size of each model. The markers show the coefficient estimates, with error bars indicating the 95\% confidence intervals (CIs) for each estimate.}
\label{gender_inferences_Likelihood_CareerLength}
\end{figure}

To ensure the robustness of our findings, we also reanalysed the data using a stricter $95\%$ probability threshold for gender assignment. Despite the reduced sample size, the main findings and conclusions remain consistent with those obtained using the $66\%$ threshold. Additionally, we performed a separate analysis using gender data derived from Genderize and Gender API. The results from this independent method were largely consistent, with one exception: the interaction term between gender and tie strength was not significant(see Figure~\ref{gender_inferences_Likelihood_CareerLength}).

\section{Adjusted collaboration statistics} 
\label{adjusted_collaboration}
To examine the relationship between gender and collaboration statistics while controlling for the potential confounding influence of publication volume, we employ a comprehensive statistical adjustment method for the collaboration statistics. This approach involves the use of Generalised Linear Models (GLMs) to disentangle the effects of the number of published papers from various collaboration metrics.

Initially, we construct separate models for each collaboration statistic, including network size, tie strength, clustering coefficient and co-authors per publication. 
We use a linear regression model to predict network size, tie strength and co-authors per publication because these collaboration statistics are continuous variables that can take on a wide range of values. We use a logistic regression model to predict the clustering coefficient since it ranges between 0 and 1.
In these models, each collaboration statistic serves as the dependent variable, while the number of publications functions as the independent variable. This modelling allows us to isolate the impact of publication quantity on each aspect of collaboration.
Subsequently, we extract the residuals from each model. These residuals represent the portion of collaboration statistics not explained by a linear relationship with the publication volume, thus effectively adjusting the collaboration statistics for the number of published papers. To facilitate comparability and eliminate scale discrepancies across different collaboration metrics, we standardise these residuals, ensuring they conform to a distribution with a mean of zero and a standard deviation of one.

\section{Gender differences in adjusted collaboration statistics}
\label{GLM_adjusted}
Examining the GLM coefficients also identifies some statistically significant differences in the mean collaboration statistics between genders. Table~\ref{GLM_gender} indicates that female authors exhibit significantly higher adjusted tie strength ($\beta = 0.173, p < 0.01$), clustering coefficient ($\beta = 0.154, p < 0.01$) and co-authors per publication ($\beta = 0.248, p < 0.01$) compared to their male counterparts. These findings indicate that female authors tend to engage in more frequent and interconnected collaborations, as well as larger collaborative groups, than male authors.

\begin{table}[H]
\centering 
\caption{Regression coefficients indicate a significant relationship between gender and adjusted collaborative behaviour statistics. Each column represents a distinct regression model. The values in parentheses below each coefficient denote the standard errors. Male is reference gender.}  
\scalebox{0.8}{
\begin{tabular}{@{\extracolsep{5pt}}l c c c c} 
\\[-1.8ex]\hline 
\hline \\[-1.8ex] 
 & \multicolumn{4}{c}{\textit{Dependent variable:}} \\ 
\cline{2-5} 
\\[-1.8ex]  & \multicolumn{1}{c}{adj. network size} & \multicolumn{1}{c}{adj. tie strength} & \multicolumn{1}{c}{adj. clustering coefficient} & \multicolumn{1}{c}{adj. co-authors per publication} \\ 
\\[-1.8ex] &  \multicolumn{1}{c}{(1)} & \multicolumn{1}{c}{(2)} & \multicolumn{1}{c}{(3)} & \multicolumn{1}{c}{(4)}\\ 
\hline \\[-1.8ex] 
Female  & 0.049$^{*}$ & 0.173$^{***}$ & 0.154$^{***}$ & 0.248$^{***}$ \\ 
        & (0.027)     & (0.027)       & (0.027)       & (0.027)       \\ 
Constant & $-$0.004    & $-$0.013$^{***}$ & $-$0.012      & $-$0.019$^{***}$ \\ 
         & (0.008)     & (0.008)       & (0.008)       & (0.008)       \\ 
\hline \\[-1.8ex] 
\textit{Note:} & \multicolumn{4}{r}{$^{*}$p$<$0.1; $^{**}$p$<$0.05; $^{***}$p$<$0.01} \\ 
\end{tabular}
}
\label{GLM_gender}
\end{table}

\pagebreak
\section{Identification of principal investigator}
\label{identify_PI}
To evaluate the impact of alphabetically-ordered author lists on the progression of authors to principal investigator (PI) status, we classify all papers into groups based on the number of authors. For each group, we quantify the number of papers arranged in alphabetical order and generate an expected number of alphabetically ordered papers, based on random chance (Table~\ref{Alphabetically_Ordered}). 
To determine the expected number of alphabetically ordered papers, we first determine the total possible permutations for $ n $ authors, denoted as $ n! $ (n factorial). Given that only one of these permutations results in an alphabetical sequence, the probability of a random alphabetical ordering is $ \frac{1}{n!} $. Consequently, by multiplying this probability with the total number of papers for each author count, we derive the expected number of papers that would be alphabetically arranged due to randomness.
Using a t-test, we compared the observed counts of papers where authors are listed in alphabetical order with the expected counts based on random chance. We found a significant difference between these counts. This result indicates that the author order in our dataset is not random and suggests a preference for alphabetical order.

\begin{table}[H]
\centering
\caption{Summary of Alphabetically-Ordered Author Lists. \textit{Authors} denotes the total number of contributing authors for a given publication, it serves as a categorical variable to group papers based on their respective authorship volume; \textit{Papers} represents the number of papers under each specific author count category; \textit{Alphabetically Ordered} quantifies the subset of publications, within each author count category where the authors are listed in strict alphabetical order; \textit{Expected randomly alphabetical} predicts the number of papers that would, by mere chance, have their authors listed alphabetically.}
\scalebox{1.1}{\begin{tabular}{ccccc}
\toprule
Authors & Papers & Alphabetically ordered & Expected randomly alphabetical \\
\midrule
2 & 52467 & 32836 & 26234 \\
3 & 44626 & 15588 & 7438 \\
4 & 28809  & 5392 & 1201  \\
5 & 17351  & 1690  & 145   \\
6 & 11376  & 630  & 16    \\
7 & 7502  & 316   & 1.49    \\
8 & 5338  & 180   & 0.13   \\
9 & 3553   & 95   & 0.01  \\
\bottomrule
\end{tabular}}
\label{Alphabetically_Ordered}
\end{table}

\begin{figure}[H]
\centering
\includegraphics[width=1\textwidth]{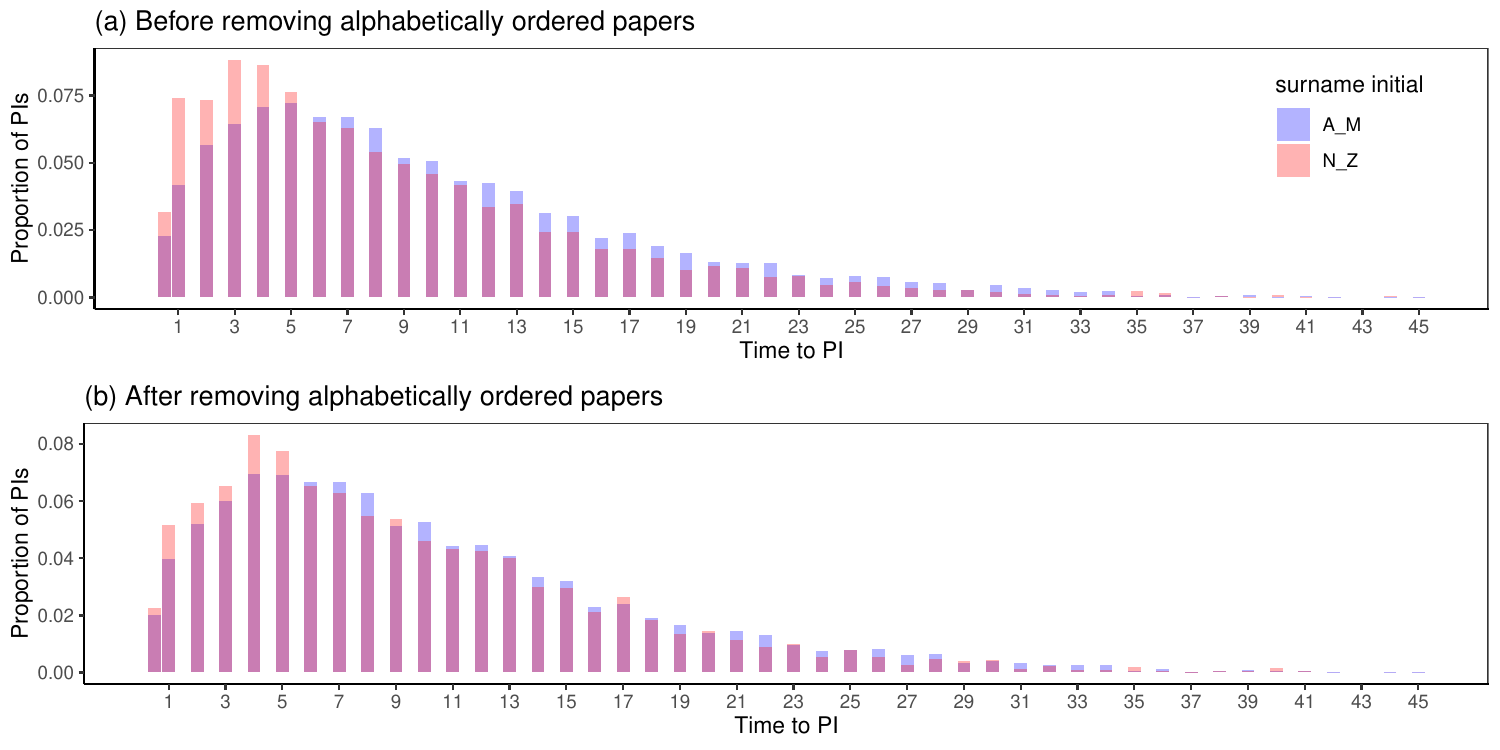}  
\caption{\textbf{Time to Principal Investigator (PI) before and after adjusting for alphabetically ordered papers.} Subfigure (A) illustrates the distribution of Time to PI across two surname-based groups (A-M and N-Z) under the initial assumption that all papers are non-alphabetically ordered. In contrast, Subfigure (B) demonstrates the altered distribution after removing papers with at least three authors that are alphabetically ordered. The assessment aims to discern the impact of surname positioning in author listings on the acceleration to PI status, revealing an initial discrepancy between Group A-M and Group N-Z, which converges to comparable distributions upon the exclusion of alphabetically ordered papers.}
\label{adjuseted_alphabetically}
\end{figure}

To assess the impact of alphabetically ordered papers on the timing of becoming a PI, we divide PIs into two groups based on the first letter of their surname: Group A-M and Group N-Z. If Time to PI is not influenced by the initial letter of the surname; then ideally the distribution of Time to PI should be similar for both groups. 
We conducted Wilcoxon signed-rank tests to assess whether the median Time to PI differed significantly between the two alphabetical groups (A-M and N-Z), both before and after accounting for alphabetically ordered papers. The results showed a significant median difference between the two groups before adjustment (p = 0.0097) but not after adjustment (p = 0.1792).
We also compare the distribution of Time to PI between these groups, both before and after removing alphabetically ordered papers. We find a discrepancy between Group A-M and Group N-Z in terms of Time to PI. 
Figure~\ref{adjuseted_alphabetically} shows PIs with surnames towards the end of the alphabet appear to reach PI status earlier before removing alphabetically ordered papers, which could be attributed to them being listed last in alphabetically ordered papers. After removing alphabetically ordered papers (alphabetically ordered papers have at least 3 authors) from consideration, the distributions of Time to PI in both groups become similar.

\section{Relationships between collaboration behaviour at PI transition and career progression}
\label{collaboration_career_PI}
\begin{figure}[H]
\centering
\includegraphics[width=1\textwidth]{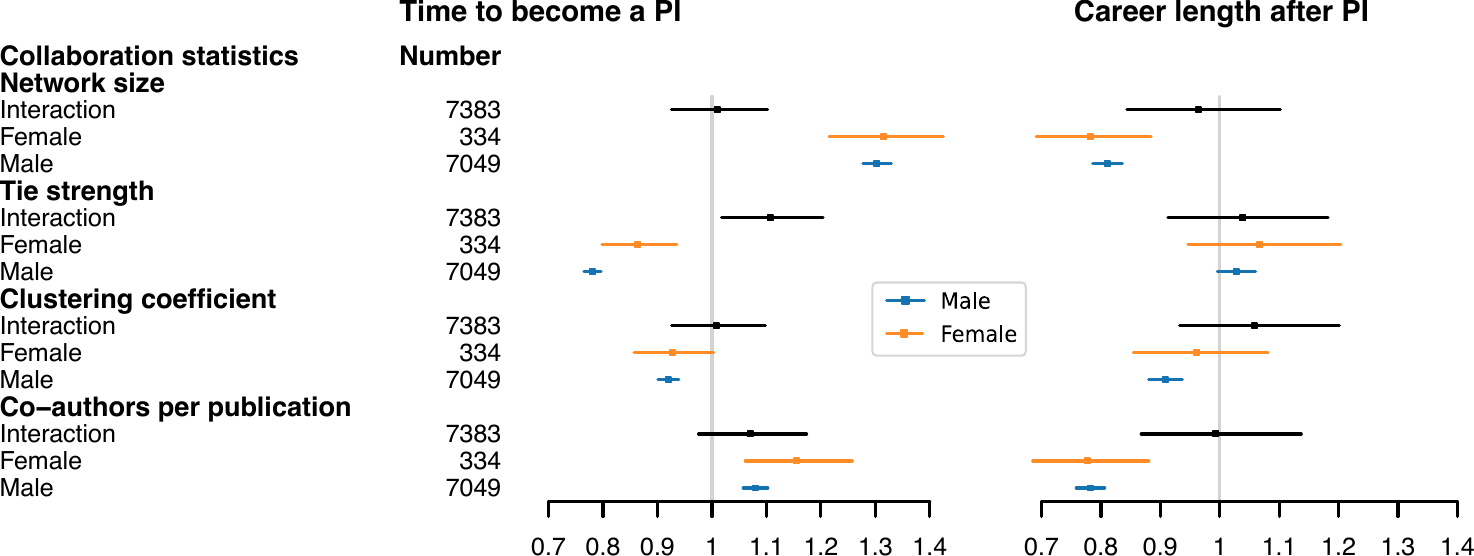} 
\caption{\textbf{The forest plots of collaboration statistics at time of becoming a PI on Time to become a PI and Career length after PI.} In each forest plot, we present three types of model for each type of collaboration statistic (network size, tie strength, clustering coefficient and co-authors per publication) at time of becoming a PI.
The \textit{interaction} model refers to the full model described in Eq.~\eqref{eq_surreg} and includes the interaction effect of gender with the collaboration statistic. The \textit{female} and \textit{male} models assess the effect of the respective collaboration statistic at time of becoming a PI on time to become a PI/career length after PI for the separate groups of female and male authors. The \textit{Number} column shows the sample size of each model. The markers show the coefficient estimates, with error bars indicating the 95\% confidence intervals (CIs) for each estimate.}
\label{PostPI_PI}
\end{figure}

We conducted additional analyses to explore the relationship between collaboration statistics at the point of becoming a Principal Investigator (PI) and career progression. This approach allows us to examine the potential influence of network structure during a key career milestone on later outcomes. Specifically, we assessed how network size, tie strength, clustering coefficient and average number of co-authors per publication at PI impact career progression.
Our findings reveal that while most conclusions from the original analysis remain consistent, there are notable exceptions. Figure~\ref{PostPI_PI} shows a larger network size at the PI stage is negatively correlated with career length post-PI, suggesting that while broad collaborations may facilitate becoming a PI, they could have adverse effects on sustaining a longer career afterward. Similarly, a higher clustering coefficient and a greater average number of co-authors per publication at PI are also negatively associated with career length after PI. These results highlight that tightly-knit or larger-scale collaborations at the PI stage might limit opportunities for sustained academic development.

\section{Journal-based author categorisation and impact analysis}
\label{Journal_based_analysis}
To control for the influence of different journals, we categorised authors based on their publication patterns across journals, dividing them into topical and prestigious journals. The topical journals include PhysRevA, PhysRevB, PhysRevC, PhysRevD, PhysRevE and the mixed category. Prestigious journals include PhysRevLett. We excluded PhysRev as it is neither a topical nor a prestigious journal and RevModPhys as it primarily publishes review articles. Journals with low representation of focal authors or those established recently were excluded which includes PhysRevApplied, PhysRevSTAB, PhysRevSTPER, PhysRevSeriesI and PhysRevX. Additionally, PhysRevAccelBeams, PhysRevFluids, PhysRevMaterials, PhysRevPhysEducRes, PhysRevResearch and PRXQuantum were excluded because they contained no focal authors in our dataset. Authors were assigned to a specific topical journal if more than 50\% of their publications were in that journal. If no journal surpassed 50\%, they were placed in the mixed category. 
For prestigious journals, we identified authors based on annual publication counts to address potential bias favouring more recent authors. Specifically, we calculated the total publications for each author annually, ranked authors by publication counts and identified the top 20\% as high-prestige authors for each year.

In topical journals, as shown in Figure~\ref{topical_journal_collaboration_differences}, we observe gender-specific differences in collaboration behaviour. Despite these differences, we find that collaboration statistics consistently impact career progression across various topical journals, underscoring their significance in the entire field of physics. For example, Figure~\ref{topical_Likelihood} illustrates that the relationship between collaboration statistics and the likelihood of becoming a PI remains consistent across these topical journals.
In prestigious journals, we find that female authors have a significantly larger network size and a lower clustering coefficient compared to the overall average for women (Figure~\ref{prestige_journal_collaboration_differences}). Additionally, they are more likely to become principal investigators (PIs) and have longer career length (Figure~\ref{prestige_journal_career_differences}). These findings align with the overall trend we observe in the relationship between collaboration statistics and career progression: those who ultimately achieve PI status tend to have larger network sizes, whereas a lower clustering coefficient is more associated with longer career lengths.

\begin{figure}[H]
\centering
\includegraphics[width=0.72\textwidth]{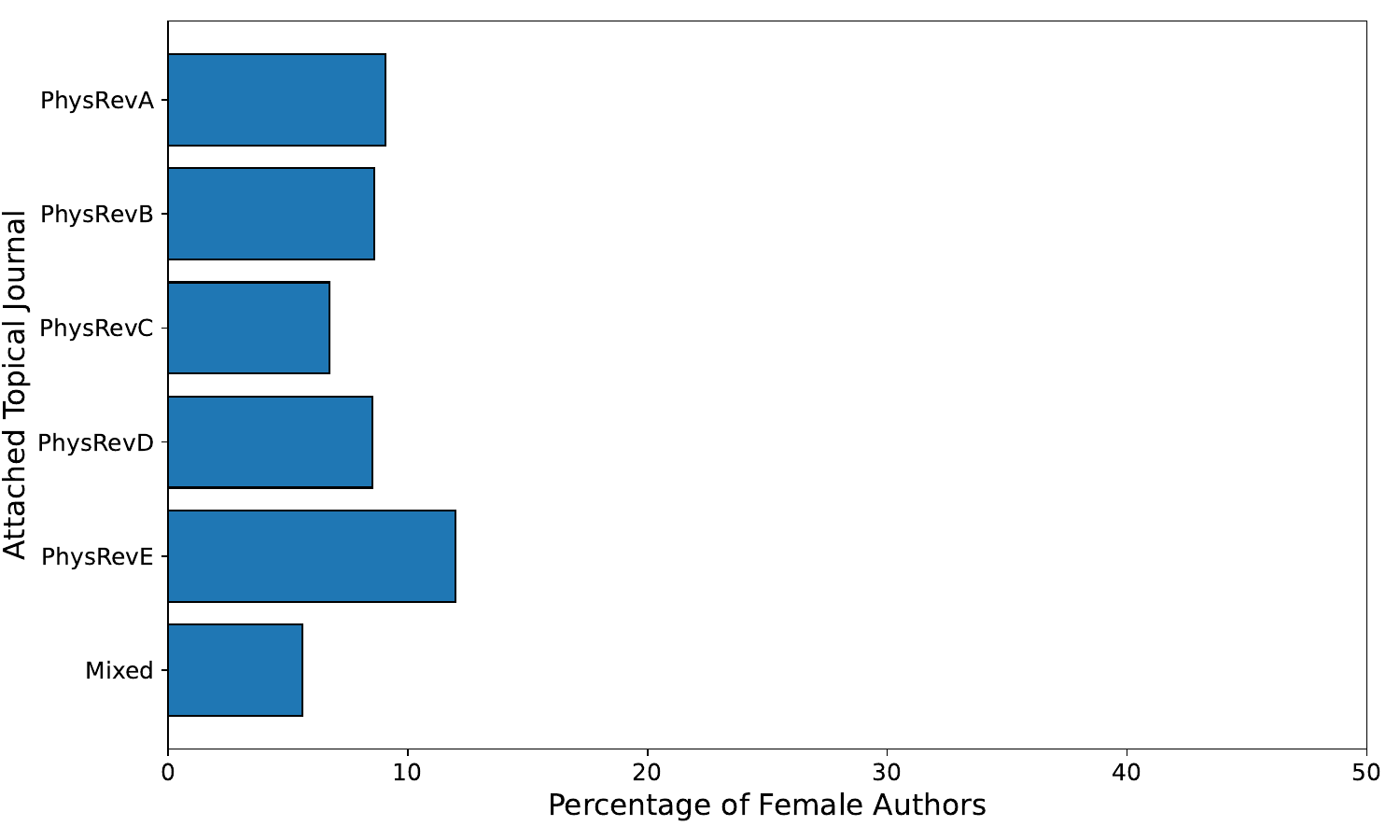}  
\caption{\textbf{Percentage of female authors by attached topical journal.} The horizontal axis represents the percentage (0\% to 50\%) and the vertical axis lists the journals.}
\label{proportion_females_topical}
\end{figure}
\begin{figure}[H]
\centering
\includegraphics[width=0.9\textwidth]{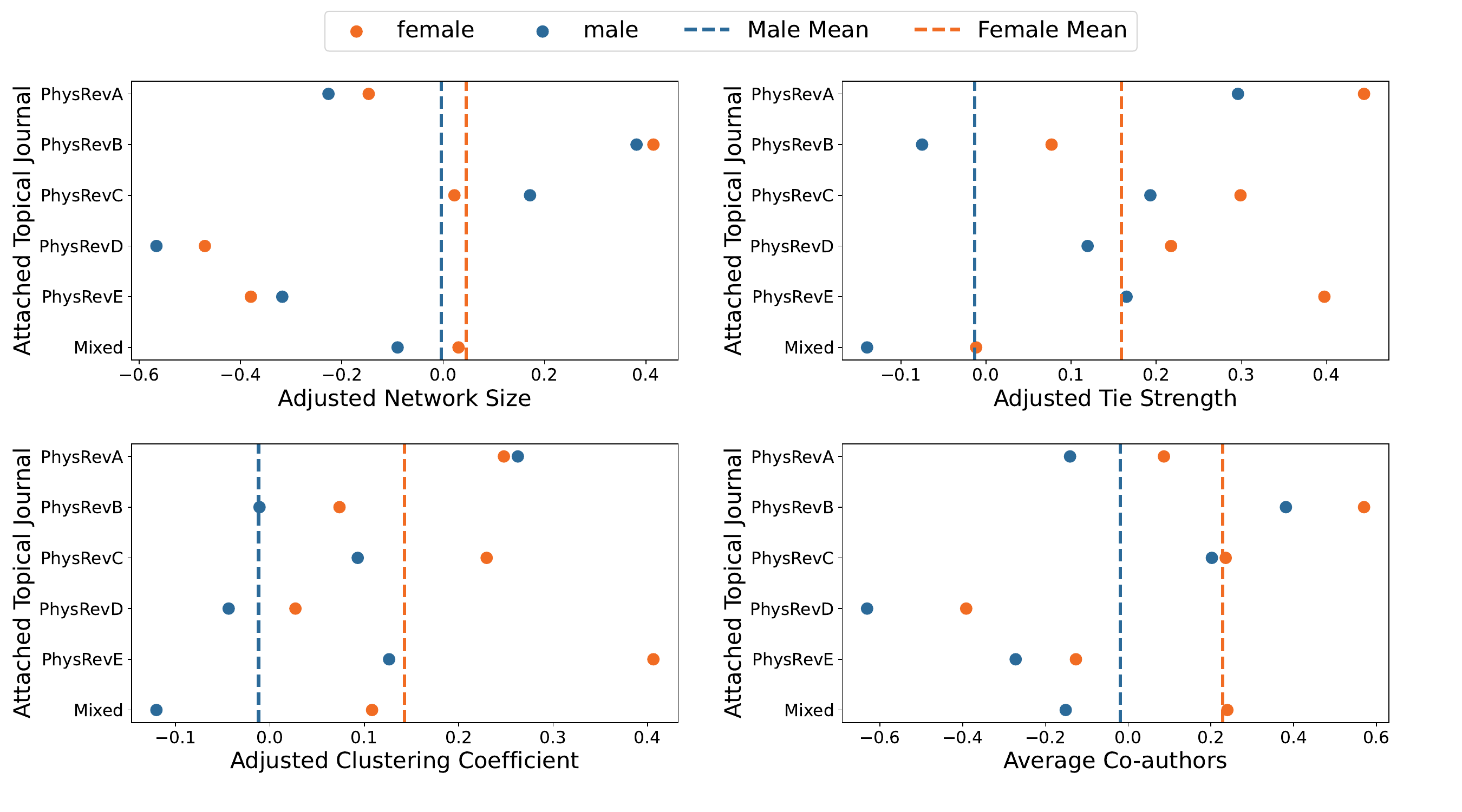}  
\caption{\textbf{Gender differences in collaboration statistics across different topical journals.} This figure shows gender differences in collaboration metrics across topical journals. Each scatterplot represents one metric, with data points showing the average values for male and female authors within each journal. The vertical dashed lines mark the overall mean values for male (blue) and female (orange) authors.}
\label{topical_journal_collaboration_differences}
\end{figure}

\begin{figure}[H]
\centering
\includegraphics[width=1\textwidth]{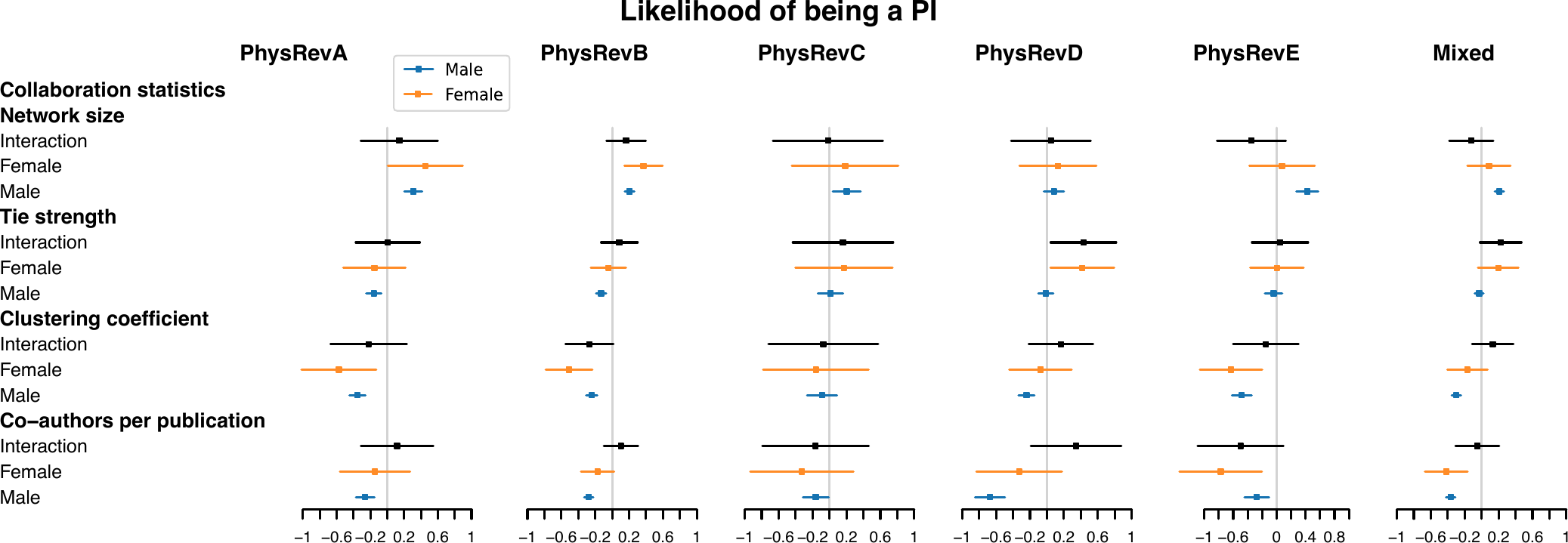}  
\caption{\textbf{The forest plot of adjusted collaboration statistics on likelihood of being a PI in different topical journals.} We present three types of model for each type of collaboration statistic (network size, tie strength, clustering coefficient and co-authors per publication). The \textit{interaction} model refers to the full model described in Eq.~\eqref{eq_logreg} and includes the interaction effect of gender with the collaboration statistic. The \textit{female} and \textit{male} models assess the effect of the respective collaboration statistic on the likelihood of being a principal investigator (PI) for the separate groups of female and male authors. The \textit{Number} column shows the sample size of each model. The markers show the coefficient estimates, with error bars indicating the 95\% confidence intervals (CIs) for each estimate.}
\label{topical_Likelihood}
\end{figure}
\begin{figure}[H]
\centering\includegraphics[width=5in]{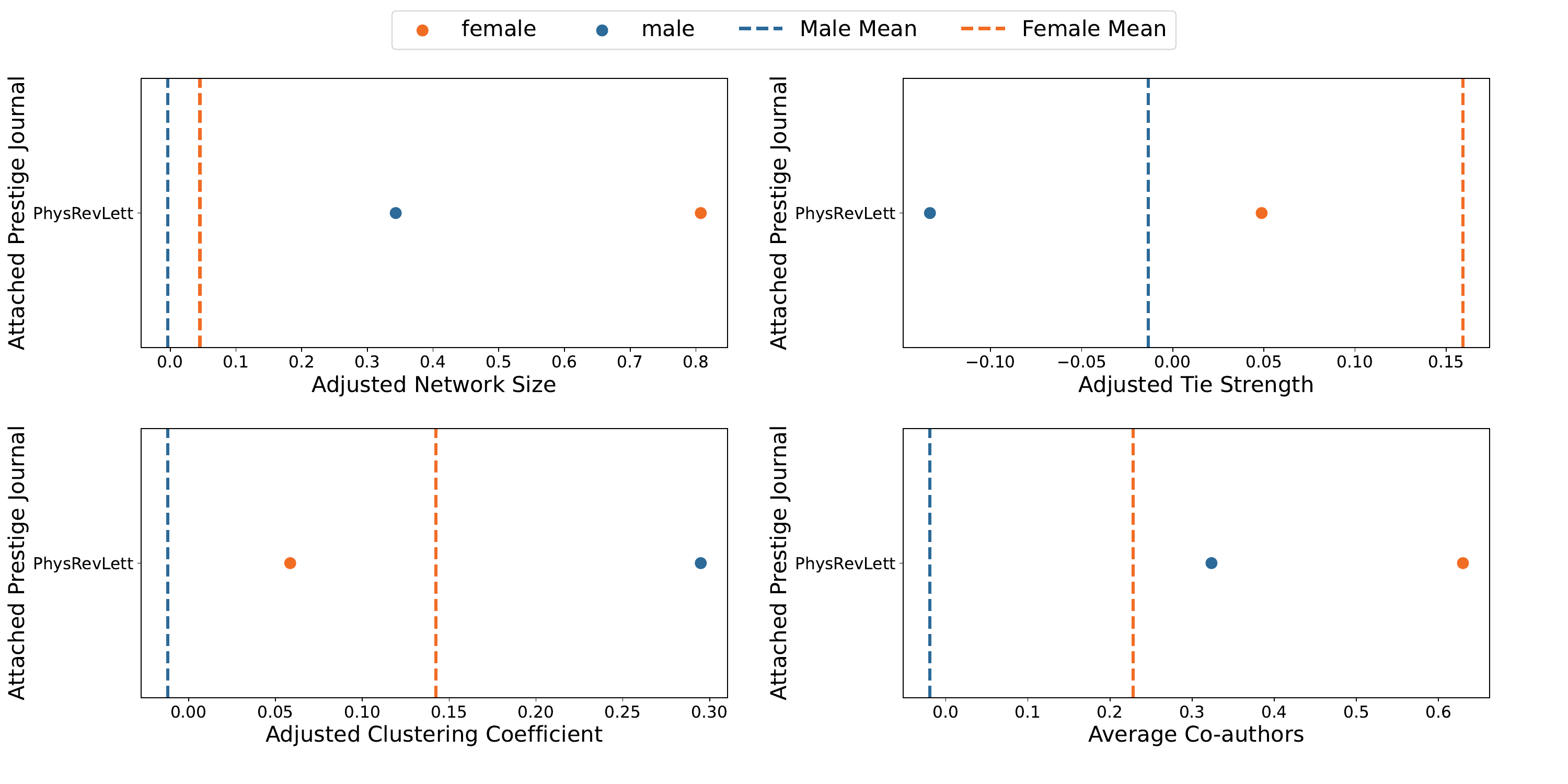}  
\caption{\textbf{Gender differences in collaboration statistics in Physical Review Letters.} This figure shows gender differences in collaboration metrics in the prestigious journal. Each scatterplot represents one metric, with data points showing the average values for male and female authors within Physical Review Letters. The vertical dashed lines mark the overall mean values for male (blue) and female (orange) authors.}
\label{prestige_journal_collaboration_differences}
\end{figure}
\unskip

\begin{figure}[H]
\centering
\includegraphics[width=0.9\textwidth]{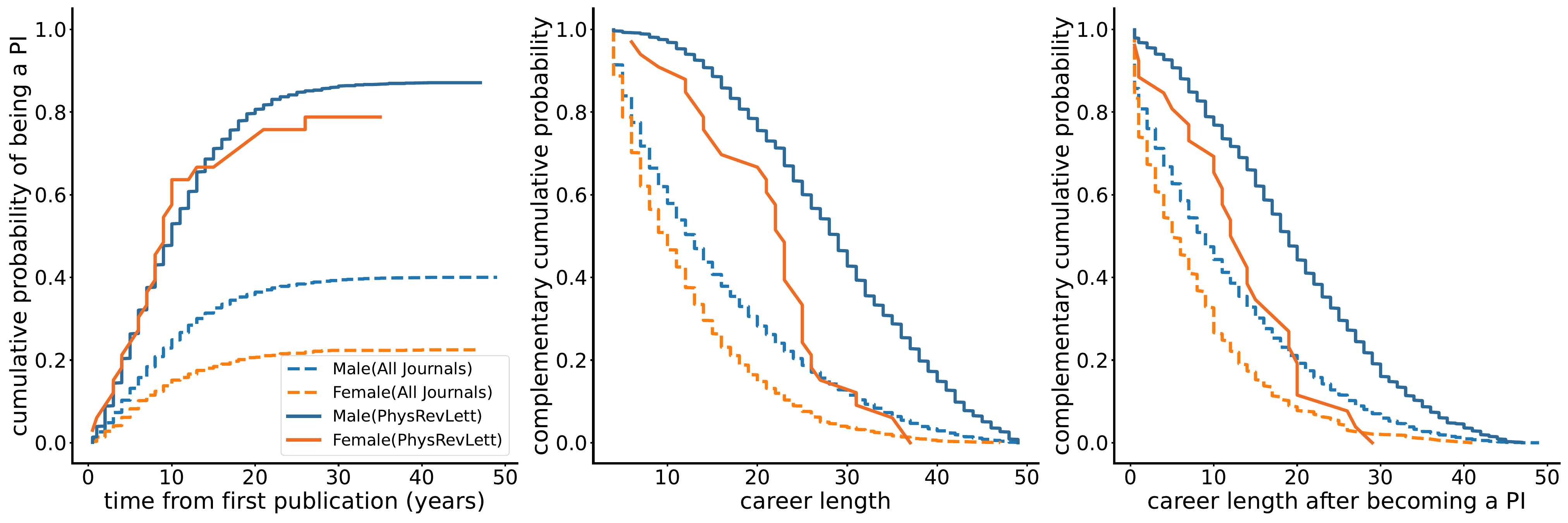}  
\caption{\textbf{Gender differences in career progression in Physical Review Letters.} This figure shows gender differences in career progression in the prestige journal. The dashed lines represent overall trends across all journals, while the solid lines focus on Physical Review Letters.}
\label{prestige_journal_career_differences}
\end{figure}

\section{Similarity in covarying career outcomes with respect to collaboration frequency}
\label{pairwise_similarity_section}
To examine whether co-authors who collaborate more frequently tend to have more similar career trajectories, we analysed pairwise similarity of how co-authors covary according to the three career outcomes: career length, time to becoming a PI and career length after becoming a PI. For each co-author pair, we first standardised their career outcomes by subtracting the mean and dividing by the standard deviation of all authors with the same number of co-authored papers. 
We then calcuate the similarity score as the product of these standardised values to obtain a similarity score for the pair. This similarity score reflects the pairwise contribution to the assortativity such that the mean of these similarity scores is equal to the assortativity coefficients in Figure~\ref{assort_career}. This approach ensures that, for each set of co-author pairs with the same number of shared publications, the average similarity score matches the Pearson correlation computed within that group.

\begin{figure}[H]
    \centering
    \includegraphics[width=\textwidth]{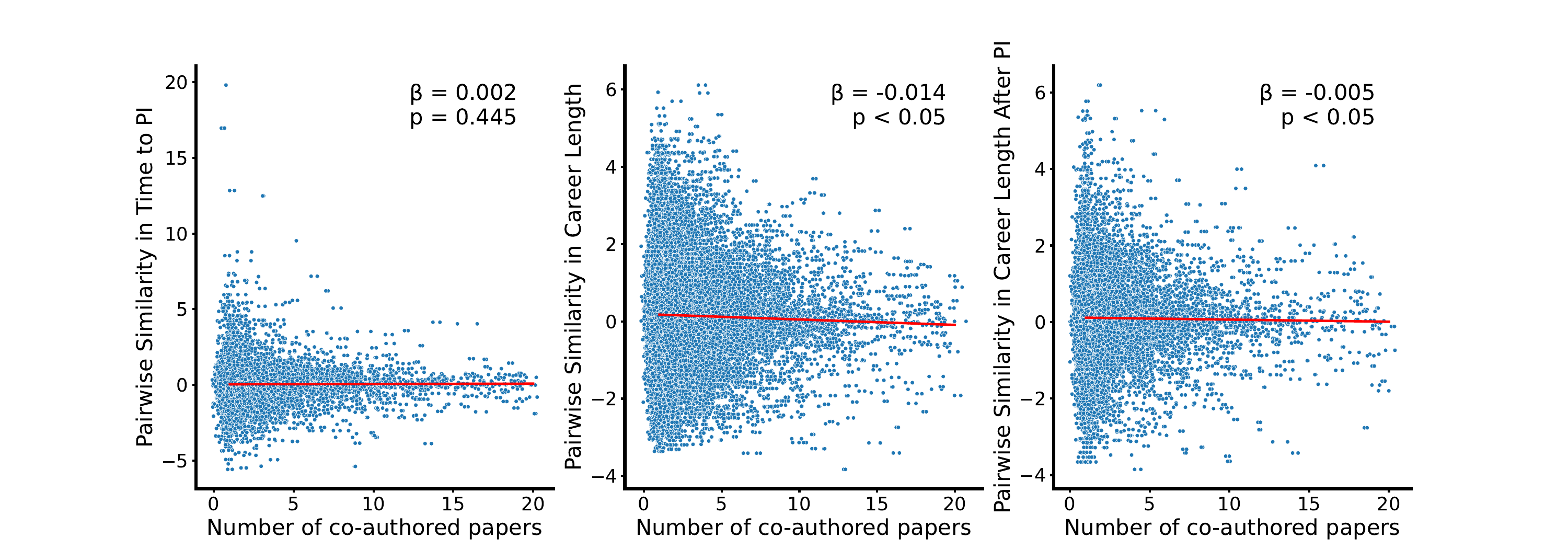}
    \caption{\textbf{Pairwise normalised similarity against co-authorship frequency.}
    Each panel shows the relationship between the number of co-authored papers and normalised pairwise similarity in time to PI, career length and career length after PI.
    Each blue point represents a co-author pair. The x-axis indicates the number of times the two authors published together. 
    Jitter is added to the x-axis to reduce overplotting.
    The y-axis represents the standardised product of each author pair's career outcome values, standardised using the mean and standard deviation within their co-authorship frequency group. 
    This value reflects the pair's contribution to the within-group Pearson correlation: positive values indicate similar career trajectories, while negative values indicate divergence.
    Red lines show linear regression fits. }
    \label{pairwise_similarity}
\end{figure}

Figure~\ref{pairwise_similarity} displays all pairwise similarity scores plotted against the number of co-authored papers, along with fitted linear regression lines to visualise potential trends. 
The results show a small but statistically significant negative association between collaboration frequency and similarity in overall career length (slope = $-0.014$, p < 0.05) and career length after PI (slope = $-0.005$, p < 0.05), indicating a very modest effect size. This suggests that frequent co-authors tend to follow slightly less similar career lengths and career length after PI. In contrast, for time to PI, no clear trend was observed. These findings indicate that repeated collaboration does not necessarily imply similar career progression.

\section{Gender composition of ego networks}
\label{female_ratio}
We additionally examined the gender composition of ego networks, focusing on both gender differences in collaboration patterns and the potential relationship between gender composition of co-authors and career progression.
For each focal author, we defined the female ratio in their ego network as the proportion of their co-authors with known gender that are female. 
We conducted a statistical test and observed a small but statistically significant negative relationship between the number of publications and the female ratio, indicating that authors with more publications tend to have slightly lower proportions of female co-authors.
In the same way that we adjusted the other collaboration statistics, we regressed the raw female ratio on the total number of publications and used the standardised residuals as an adjusted measure of gender composition in the ego network.

We first compared this adjusted female ratio between male and female authors. On average, female authors tend to collaborate with a higher proportion of female co-authors than male authors. To formally assess this difference, we conducted an independent samples t-test. The results show that female authors have significantly higher values than male authors (mean for females = $0.358$, mean for males = $-0.030$, $t = -11.346$, $p < 0.001$), indicating that, even after adjusting for publication volume, female authors are more likely to form gender-homogeneous collaboration networks.

We then investigated whether the adjusted female ratio is associated with career progression outcomes. We applied logistic regression and accelerated failure time (AFT) models, including the adjusted female ratio as a covariate when predicting (i) the likelihood of becoming a PI, (ii) career length, (iii) time to becoming a PI and (iv) career length after becoming a PI.
Table~\ref{Female_ratio_coefficient} presents the regression results. The interaction terms between gender and the adjusted female ratio are not statistically significant in any of the models, suggesting that the effect of female ratio in ego network does not differ significantly between male and female authors. For male authors, a higher female ratio is significantly associated with a longer time to become a PI (time ratio = 1.034, $p < 0.01$). No consistent pattern is observed for female authors across the four outcomes.

\begin{table}[H]
\centering 
\caption{\textbf{Regression results for the effect of female ratio in ego networks on career progression.} This table shows the coefficients from logistic regression (Model 1) and accelerated failure time (AFT) models (Models 2--4) assessing the association between the adjusted female ratio in an author's ego network and four career progression outcomes: (1) likelihood of becoming a PI, (2) career length, (3) time to become a PI and (4) career length after PI. ``Interaction'' refers to the interaction term between the female ratio and gender. ``Female'' and ``Male'' rows represent gender-specific models. Coefficients are odds ratios (Model 1) or time ratios (Models 2--4), with 95\% confidence intervals in parentheses.}

\scalebox{0.8}{
\begin{tabular}{@{\extracolsep{5pt}}l c c c c} 
\\[-1.8ex]\hline 
\hline \\[-1.8ex] 
 & \multicolumn{4}{c}{\textit{Dependent variable:}} \\ 
\cline{2-5} 
\\[-1.8ex] & \multicolumn{1}{c}{likelihood of being a PI} & \multicolumn{1}{c}{career length} & \multicolumn{1}{c}{time to become a PI} & \multicolumn{1}{c}{career length after PI} \\ 
\\[-1.8ex] & \multicolumn{1}{c}{(1)} & \multicolumn{1}{c}{(2)} & \multicolumn{1}{c}{(3)} & \multicolumn{1}{c}{(4)}\\ 
\hline \\[-1.8ex] 
Interaction  & $-$0.033 & 0.997 & 1.006 & 0.939 \\ 
             & ($-$0.133,\ 0.067) & (0.968,\ 1.026) & (0.931,\ 1.086) & (0.838,\ 1.052) \\ 
Female       & $-$0.009 & 0.997 & 1.040 & 0.918 \\ 
             & ($-$0.105,\ 0.086) & (0.973,\ 1.022) & (0.970,\ 1.115) & (0.828,\ 1.018) \\ 
Male         & 0.024 & 1.000 & 1.034$^{***}$ & 0.978 \\ 
             & ($-$0.007,\ 0.055) & (0.990,\ 1.011) & (1.010,\ 1.059) & (0.944,\ 1.013) \\ 
\hline \\[-1.8ex] 
\textit{Note:} & \multicolumn{4}{r}{$^{*}$p$<$0.1;\ $^{**}$p$<$0.05;\ $^{***}$p$<$0.01} \\ 
\end{tabular}
}
\label{Female_ratio_coefficient}
\end{table}

\end{document}